\documentclass[useAMS, usenatbib, usedcolumn, fleqn]{mnras}
\usepackage{graphicx}
\usepackage{amsmath, amssymb, blindtext}

\usepackage{times}
\def\sun{\hbox{$\odot$}}
\newcommand{\mamo}[1]{\mbox{$#1$}}
\newcommand{\unit}[1]{\ifmmode \:\mbox{\rm #1}\else \mbox{#1}\fi}
\newcommand{\sbr}[1]{_{\textnormal{#1}}}

\newcommand{\ten}[1]{\mamo{\times 10^{#1}}}

\newcommand{\secref}[1]{Section~\ref{sec:#1}}

\newcommand{\equref}[1]{equation~(\ref{eq:#1})}

\newcommand{\figref}[1]{Fig.~\ref{fig:#1}}
\newcommand{\tabref}[1]{Table~\ref{tab:#1}}

\newcommand{\sersic}{S\'{e}rsic }
\newcommand{\ratio}{$R\sbr{e}/R_{200}$}
\newcommand{\fr}{\mamo{r_{1/2}}}

\begin{document}

\title[Sizes of globular cluster systems]{The correlation between the sizes of globular cluster systems and their host dark matter haloes}

\author[]{\parbox{\textwidth}{
\vspace{-0.5cm}
\raggedright
\mbox{Michael J. Hudson\unskip$^{1,2}$ and}
\mbox{Bailey Robison$^{1}$}
}
\vspace{0.4cm}\\
\parbox{\textwidth}{
$^1$ Department of Physics \& Astronomy, University of Waterloo,
Waterloo, ON, N2L 3G1, Canada.\\
$^2$ Perimeter Institute for Theoretical Physics, 31 Caroline St. N.,
Waterloo, ON, N2L 2Y5, Canada.\\
\vspace{-0.5cm}}
}

\maketitle

\begin{abstract}

The sizes of entire systems of globular clusters (GCs) depend on the formation and destruction histories of the GCs themselves, but also on the assembly, merger and accretion history of the dark matter (DM) haloes that they inhabit. Recent work has shown a linear relation between total mass of globular clusters in the globular cluster system and the mass of its host dark matter halo, calibrated from weak lensing. Here we extend this to GC system sizes, by studying the radial density profiles of GCs around galaxies in nearby galaxy groups. 
We find that radial density profiles of the GC systems are well fit with a de Vaucouleurs profile. Combining our results with those from the literature, we find tight relationship ($\sim 0.2$ dex scatter) between the effective radius of the GC system and the virial radius (or mass) of its host DM halo. The steep non-linear dependence of this relationship ($R\sbr{e, GCS} \propto R_{200}^{2.5 - 3}$) is currently not well understood, but is an important clue regarding the assembly history of DM haloes and of the GC systems that they host.

\end{abstract}

\begin{keywords}
galaxies: groups -- galaxies: haloes -- globular clusters \vspace{-0.55cm}
\end{keywords}

\section{Introduction}
Globular clusters (GCs) trace the formation and evolution of galaxies and of their dark matter (DM) haloes. Recent studies have shown that the mass of the entire GC system (GCS) is correlated linearly with the \emph{DM} mass of the host halo \citep{Hudson14, Harris15, ForAlaRom16}, extending and updating previous work (\citealt{BlaTonMet97, SpiFor09}; see \citealt{Harris15} and references therein). This correlation is surprising, because the mass in \emph{stars} in the host galaxy is not linearly related to its DM halo mass, but rather has a break around the luminosity of an $\sim L_{*}$ galaxy, or equivalently, a  $\sim 10^{12} M_{\odot}$ halo mass \citep{MarHud02, BehWecCon13}. The linear correlation between GCS mass and DM halo mass may reflect the early formation epoch of GCs, before feedback regulated the formation of stars in the host galaxy.

The GCS of a halo will be affected by the accretion and tidal stripping of its satellite galaxies, as first suggested by \cite{SeaZin78} for the Milky Way halo, and subsequently updated to galaxy group and cluster-scale haloes in the context of hierarchical cosmological models \citep{WesCotJon95, Forbes97, CotMarWes98, BeaBauFor02, Ton13}. When a satellite galaxy falls into a larger halo, its DM is stripped and becomes part of the host halo. A statistical detection of this has been observed via weak lensing in galaxy clusters \citep{LimKneBar07, NatKneSma09, LiShaKne16} and in galaxy groups \citep{GilHudErb13, LiShaMo14}. Tidal stripping is most likely to affect the least bound objects, and so, because the GC system as a whole is less spatially extended than the DM, it is likely to be stripped to a lesser degree than the DM \citep{YahBek05, SmiSanFel13, SmiSanBea15, RamCoeMur15}. Nevertheless, some of a satellite's GCs may be stripped from the satellite's halo, in which case these GCs will join the GCS of the host dark matter halo. There is possible evidence of tidal stripping of GC systems, in the form of arcs or tails of GCs around satellite galaxies in clusters \citep{%
RomStrBro12,%
BloForBro12,%
ChoBlaChi16,%
VogHilRic16%
}
as well as around galaxies such as M31 \citep{MacHuxFer10}.
The indirect evidence for tidal stripping is also quite strong. First, recent studies have uncovered large, spatially-extended populations of mostly blue/metal-poor intracluster GCs in nearby galaxy clusters such as Coma \citep{Peng11} and Virgo \citep{Lee10, Durrell14}.
Second, a number of studies have found evidence that satellite galaxies in clusters have lower GC specific frequencies than dominant central galaxies \citep{FleHarPri95, Forbes97, PenJorCot08, WehHarWhi08, CoeMurDon09}.

After cluster pericentric passage, the ``backsplash'' orbit of a satellite galaxy may reach an apocentre distance of more than twice the virial radius \citep{BalNavMor00, GilKneGib05, LudNavSpr09, OmaHudBeh13}. Objects that are tidally stripped from the satellite -- such as GCs -- will follow approximately the same orbit as the satellite galaxy but either leading or lagging. As a result, one may also expect to find tidally-stripped GCs at large clustercentric radii. To detect this population, imaging should extend beyond the virial radius of the host halo, which may be as much as 2 Mpc for a rich galaxy cluster.

Accretion and tidal stripping of satellites is also expected to occur in lower mass haloes such as those hosting galaxy groups, but to date there have been no detections of ``intra-group'' GCs, including in the Local Group \citep{MacBeaLea16}. Indeed galaxy groups, with halo masses $\sim 10^{13} M_{\sun}$, contribute the most of all DM haloes to the overall abundance of GCs in the Universe \citep{Har16}.  The primary goal of this paper is to study the large-scale spatial distribution of the GC systems around group galaxies and within galaxy groups. The virial radius of $10^{13} M_{\sun}$ galaxy group is $\sim 500$ kpc, which at a distance of 30 Mpc  corresponds to a degree on the sky. Hence, to study the distribution of GCs on large scales, wide-field imaging is required. In recent years, a number of authors have conducted wide-field imaging of GCs reaching galactocentric distances of $\sim$ 100 kpc \citep{Rhode04, BasFaiFor06, RhoZepKun07, Har09, HarGomHar12, PotForRom13, RejHarGre14, Kartha14, Hargis14, Kartha16}

It has long been known that more luminous galaxies have a GCS with a shallower radial profiles \citep{Har86, Kissler97, AshZep98}. Early work on the spatial distribution of GCs is reviewed in \cite{BroStr06}. More recent work has compared the ``extent'' of the GCS, where extent is defined as the radius at which the GC density is consistent with zero, to the stellar mass of the host galaxy \citep{RhoZepKun07}. There have been no attempts to compare the overall spatial scale of the GCS to the size or mass of its host dark matter halo.

In this paper, we identify nearby galaxy groups within the footprint of the CFHT Legacy Survey (CFHTLS) fields and identify candidate GCs around the group galaxies. An outline of this paper is as follows: in \secref{data} we describe the data sets and sample selection used to compile samples of candidate GCs. \secref{profile} discusses the radial profiles of GCs and compares the total GC counts to previous work. In \secref{spatial}, we combine the results from this paper with other measurements from the literature, and explore the correlations between the physical size of the GCS system with the effective radius of the stellar light and the virial radius and/or mass of the host halo. We discuss the implications of these results in \secref{discuss} and conclude in \secref{conc}

\section{Data}
\label{sec:data}

\makeatletter{}\begin{table*}
\caption{Important parameters for each galaxy analyzed in this paper. The CFHTLS field and group where the galaxy can be found as well as the morphological type (from NED) are given. The $K$ magnitude, $M_K$, as well as $L_*$, $M_*$, and $M_{200}$ from \secref{galaxies} are also given.}
\begin{tabular}{lllclcccc}
\hline
Galaxy&Field&Group&Distance&Morphology&$M_K$&$L_{*}$&$M_{*}$&$M_{200}$\\
&&&(Mpc)&&&($M_{\odot}$)&($M_{\odot}$)&($M_{\odot}$)\\
\hline
IC 219&W1-0-0&IC 219&72.7&E&-24.33&1.112e+11&9.353e+10&5.020e+12\\
NGC 883&W1-0-0&''&&S0&-25.42&3.020e+11&2.509e+11&2.704e+13\\
\hline
NGC 942+943&W1+3-4&NGC 943&67.0&S0&-25.16&2.373e+11&1.972e+11&1.744e+13\\
\hline
NGC 2695&W2-0+1&NGC 2695&26.5&S0&-23.26&4.140e+10&3.461e+10&1.407e+12\\
NGC 2698&W2-0+1&''&&S0&-23.27&4.182e+10&3.475e+10&1.413e+12\\
NGC 2699&W2-0+1&''&&Sb&-22.74&2.569e+10&2.044e+10&8.624e+11\\
\hline
NGC 5473&W3-2-0&NGC 5473&26.2&S0&-23.74&6.398e+10&5.267e+10&2.267e+12\\
NGC 5475&W3-2+1&''&&Sa&-22.69&2.446e+10&1.946e+10&8.281e+11\\
NGC 5485&W3-2-0&''&&S0&-23.69&6.155e+10&5.067e+10&2.162e+12\\
\end{tabular}
\label{tab:important}
\end{table*}

\subsection{Group Selection and Host Galaxy Data}
\label{sec:galaxies}

We targeted galaxies in groups that were in CFHTLS fields.  Four groups with $cz < 5000$ km/s ($d \lesssim 70$ Mpc) that overlapped the CFHTLS-Wide footprint were selected from the 2MASS-based 2M++ group catalogue of \cite{Lavaux11}.  For comparison with the properties of its GCS, we require, for each galaxy, its distance, stellar mass, effective radius and host halo mass. These were determined as follows:

\begin{itemize}

\item Distances are based on the recession velocity from NED, adopting a value of 70\,km/s/Mpc for the Hubble parameter ($H_0$). 

\item Stellar masses are derived from 2MASS $K$-band magnitudes. We adopt a solar $K$-band magnitude of 3.28\, \citep{Binney98}, to calculate the luminosity of each galaxy in units of $L_{\sun}$.  The galaxy's stellar mass is then obtained via the $K-$band stellar-mass-to-light ratio ($M_{*}/L_{K}$) from \cite{Bell03}, using the $B-V$ colour of each galaxy from the Third Reference Catalog of Bright Galaxies \citep{RC3}. In cases where the colours were not available in the catalogue, the average colour of the galaxy's morphological type \citep{Fukugita95} was used.

\item Effective radii of the galaxy light in the $K$-band will be used in \secref{spatial}. These were kindly provided to us by J.\ Lucey (priv.\ comm.), using the methods described in \cite{Campbell14}.

\item The halo mass ($M_{200}$) and halo virial radius ($R_{200}$) are critical parameters in this study. Unfortunately, it is impossible to measure dark matter halo masses out to the virial radius directly in individual galaxies.  Therefore we use the mean relationship between dark matter halo mass and stellar mass, calibrated via weak gravitational lensing \citep{Hudson15}. Specifically, we use the fits described in Appendix~C of that paper, extrapolated to $z \sim 0$ to obtain halo masses (and hence virial radii). Note also that the intrinsic scatter in $M_{200}$ at fixed $M_{*}$ is estimated to be in the range 0.15-0.20 dex.

It is important to understand that the $M_{200}$ masses determined from this relationship refer to the mass of the \emph{entire} halo of which the galaxy is assumed to be the central member. For massive galaxies that are the dominant galaxy of a group or cluster, the halo mass is therefore the total mass of the group or cluster.  For example, the stellar mass of NGC 883 is $2.5\,\times\,10^{11}\,M_{\sun}$, typical of the dominant ``brightest group galaxies'' (hereafter BGG).  The corresponding halo mass of its group is then $2.7\,\times\,10^{13}\,M_{\odot}.$. 
\end{itemize}

Parameters of massive galaxies in groups are summarized in \tabref{important}. There is a special case in \tabref{important}: the galaxies in the pair NGC 942+943 are too close to separate their GC systems. We therefore combine their GC counts and their stellar masses and treat the pair as a single ``BGG''.

\subsection{Globular Clusters: Catalogues and Selection Criteria}
\label{sec:candidates}

\subsubsection{Catalogues}

The photometric data used to select GC candidates were obtained from the Wide component of the Canada-France-Hawaii Telescope Legacy Survey\footnote{http://www.cfht.hawaii.edu/Science/CFHTLS/} (CFHTLS). This survey used the MegaCam instrument, with a $1^{\circ} \times 1^{\circ}$ field of view, and a scale of $0^{''}$\!\!\!.187 per pixel. CFHTLS-Wide covers 155\,deg$^2$ across four patches: W1, W2, W3, and W4. Each of these patches comprises several fields with coverage in five Megacam filters: $u^*$, $g'$, $r'$, $i'$, $z'$.  The last four are similar to SDSS filters. Hereafter we drop the prime notation. The limiting magnitude in the $i$-band is $\sim 24.7$ for a point source.  Calibrated images and source catalogues were produced by Terapix and made available through the CFHTLS-T0007 release.  For more information on the calibration and other technical details, see \cite{Hudelot12}

\subsubsection{Size and magnitude selection}

In order to distinguish globular clusters from the other sources, we impose several selection criteria. To avoid contamination from bright stars, nearby bright galaxies, defects and edge effects, masks were created and applied to each image.   We then select GC candidates based on apparent $i$~magnitude, half-flux radius (\fr) and $(g-i)$ colour. 

\begin{figure}
\centering
\includegraphics[width=0.45\textwidth]{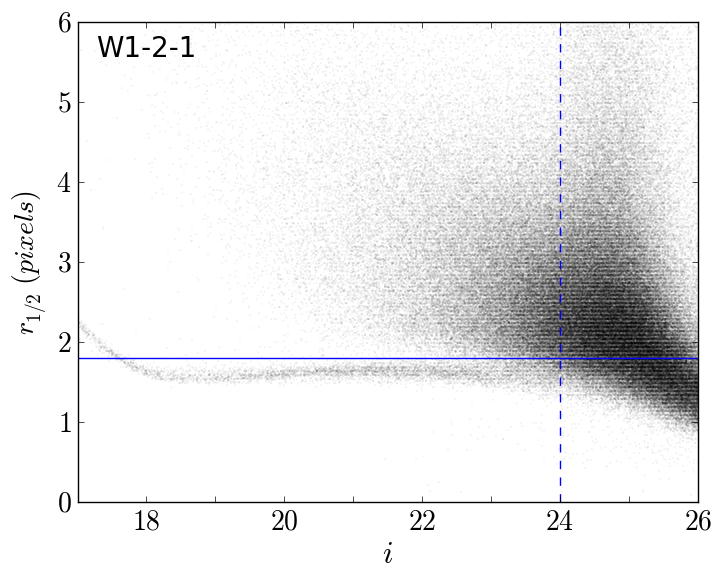}
\caption{Half-flux radius \fr\, plotted against $i$-band magnitude for the W1-2-1 field. The half flux radius value is the radius of the object (in pixels on the image) that encloses half of the object's total flux. The large region in the bottom, right corner is mostly unresolved background galaxies. To avoid selecting these, we impose a $i$ criterion of $i~<~24$, which is represented by a vertical dashed blue line. The thin locus on the left side are point sources, mostly stars but also some GCs. The adopted cut on flux radius for this field is shown by a solid blue line.}
\label{fig:fluxrad}
\end{figure}

At the distances of the galaxies in this paper, we would expect GC candidates to appear as unresolved point sources. We first select point sources by setting an upper limit on \fr. Figure~\ref{fig:fluxrad} shows a representative plot of \fr\, vs $i$ for one of the CFHTLS fields. The line in the bottom left portion of the figure are point sources, most of which are foreground stars but also including GCs.  Because the seeing varies from exposure, the cut in \fr\ used to define point sources varies from field to field. In \figref{fluxrad}, background galaxies dominate at magnitudes fainter than $i \sim 24$, so we limit GCs to  $i < 24$. This criterion is shown in \figref{fluxrad} as a vertical dashed blue line.
\begin{figure}
\centering
\includegraphics[width=0.45\textwidth]{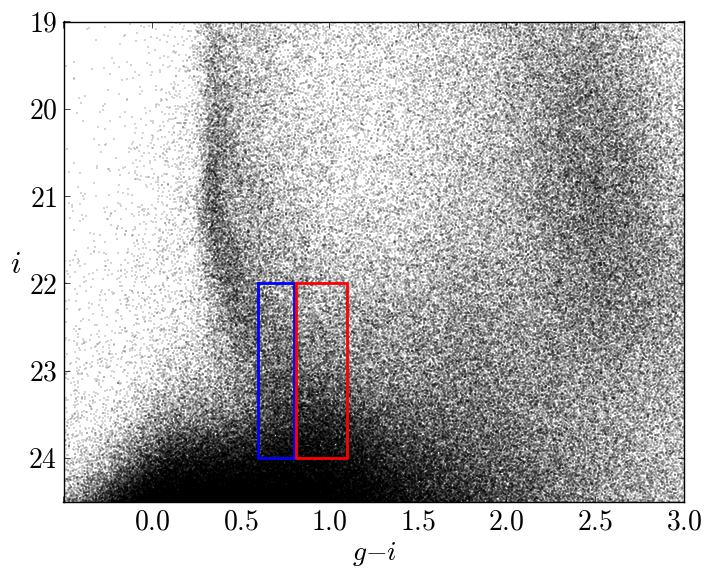}
\caption{Colour-magnitude diagram that combines point sources from every field. The dense column at $(g-i)~\sim~0.5$ is made up mostly of stars. The area at the base of the column is likely to contain globular clusters. Boxes have been placed around this region to represent our selection criteria of $i\sbr{bright} < i < 24$ and $0.6 < (g-i) < 1.1$. Although $i\sbr{bright}$ takes a different value for each galaxy, in the figure the box extends upwards to $i = 22$, a typical value of $i\sbr{bright}$. The red box surrounds the region of red GCs and the blue box surrounds the region of blue GCs. The distinction is made at $(g-i)$ colour of $0.8$.}
\label{fig:magcolor}
\end{figure}

\subsubsection{Globular Cluster Luminosity Function and Corrections for Incompleteness}

The globular cluster luminosity function (GCLF) is the number of clusters per unit magnitude, and is well described by a Gaussian whose mean and standard deviation depend on the absolute magnitude of the host galaxy. The full details of the adopted GCLFs are given in \secref{gclf}. The GCLF allows a correction for undetected GCs below the flux limit. 

If only a faint $i$-band magnitude selection limit were imposed, with no bright limit, then a large number of stars would also be selected by the criteria and treated as GC candidates.  In order to minimize the stellar contamination and to maximize the number of GCs, we use the GCLFs (\secref{gclf}) to determine the bright $i$-band magnitude limit ($i\sbr{bright}$~\textless~$i$~\textless~24) on a galaxy-by-galaxy basis.
Specifically, we select $i\sbr{bright}$ in such away that we lose no more than 10\% of the GCLF visible at $i  < 24$.

Using the GCLF we obtained for each galaxy in \secref{gclf}, we calculate the fraction of the total GCLF that lies within our observed magnitude range ($i\sbr{bright}$~\textless~$i$~\textless~24). This factor allows us to correct the raw counts for the GCs that lie outside the magnitude range.

\begin{figure}
\centering
\includegraphics[width=0.45\textwidth]{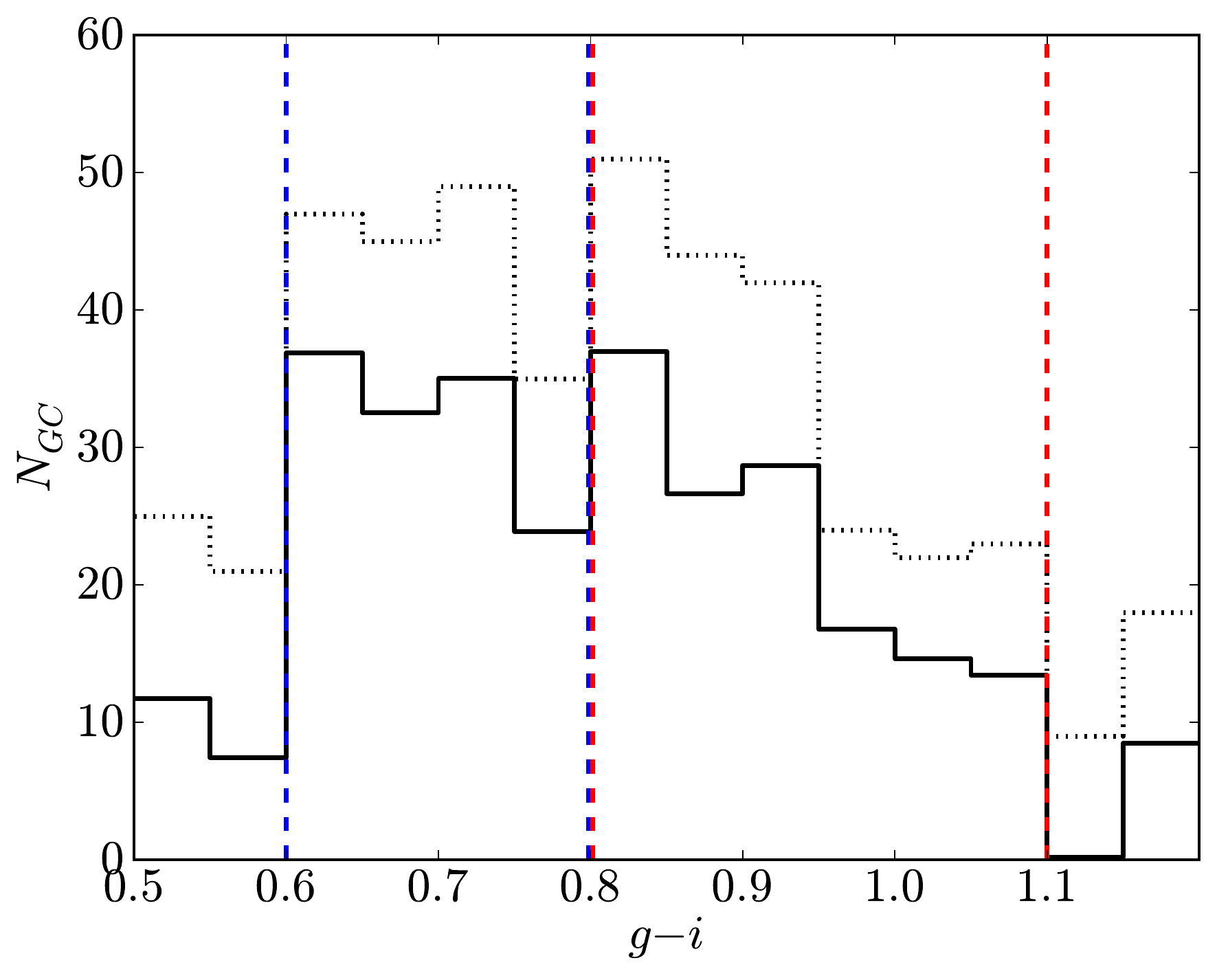}
\caption{Stacked histogram of the $(g-i)$ colours of all GCs within 3~\!$R\sbr{e}$ of all galaxies. The dotted line represents the observed histogram, while the solid black line represents the background-subtracted histogram. The dashed vertical lines demonstrate the range of colour assumed to be dominated by GCs. The blue and red dashed lines show the colours adopted to separate blue from red GCs.}
\label{fig:colorhist}
\end{figure}

\subsubsection{Colour Selection}
\label{sec:colourselect}
We determine the range of $(g-i)$ colour in which we would expect to find globular clusters by plotting a colour-magnitude diagram (CMD) of $i$ vs $(g-i)$ colour \citep{Durrell14, Escudero15, Salinas15}, shown in \figref{magcolor}. The dense column at $(g-i) = 0.5$ is made up mostly of stars. \cite{Durrell14} suggest one would expect to find GCs near this area and impose a criterion of $0.55$~\textless~$(g-i)$~\textless~$1.15$. To determine exactly what range of $(g-i)$~colour we expect to find GCs, we create a $(g-i)$ histogram of GC candidate colours near this region of colour-magnitude space. We include GC candidates that meet our selection criteria within ~0.15~\!$R_{200}$ around each galaxy, a radius which is (on average) $\sim 3$ $R\sbr{e, GCS}$. However, in every field there is background contamination which may vary as a function of colour. In order to correct for this, we subtract the background estimated from unmasked areas of the field, far from galaxies. The background-subtracted histograms of each galaxy co-added to obtain a total colour histogram of every object near the massive galaxies, as shown in \figref{colorhist}. There is a clear excess of GCs within the range $0.6$~\textless~$(g-i)$~\textless~$1.1$, which we adopt as our GC colour selection criterion.

GCs are often separated into red and blue old stellar populations with differing metallicities, so we might expect to see evidence of bimodality in the $(g-i)$ histogram. There is evidence of a possible bimodal distribution of GCs across the range of $(g-i)$ colour, however it is not clear enough to be convincing. Instead of using the histogram to determine how to distinguish between red and blue GCs, we adopt the colour separation $(g-i)$ = 0.80 used by \cite{Durrell14} to separate red and blue GCs. These criteria are displayed as boxes in \figref{magcolor}.

In summary, GCs are selected by $i$-band magnitude, within a range of $(g-i)$ color, and below a certain \fr\, limit. We use these selection criteria to determine which galaxies are likely to have a significant GC population. In total, we start with a set of several galaxies in 8 different groups. However, not all of these galaxies have a significant number of GCs.  After applying the selection criteria, we observe 9 galaxies with a significant excess of GCs above the background. The rest of this paper will focus on these 9 galaxies.

\section{GCS Density Profiles}
\label{sec:profile}
We now turn to the number and density profile of GCs around their parent galaxies. We consider common functional forms used to model the surface density of GCs and apply these models to the well-studied Milky Way GCS. We then fit the models to the group galaxies discussed in this paper.

\subsection{\sersic and Power Law Models}
By modelling GC number density as a function of galactocentric distance, it is possible to extrapolate to the number of GCs around each galaxy. GC density is often modelled with a \sersic profile and a power law \citep[e.g.][]{Rhode04,Faifer11,Kartha14}. We consider both functional forms in this paper.

A \sersic intensity profile  \citep{Sersic63,Sersic68} is often used to model the surface brightness profiles of galaxies. It is represented by the function
\begin{equation}
\label{eq:sersic}
\Sigma\sbr{S}(R) = \Sigma\sbr{e} \exp \left(-b_n\left[\left(\frac{R}{R\sbr{e}}\right)^{1/n}-1\right]\right)\,,
\end{equation}
where $\Sigma\sbr{e}$ is the surface number density of GCs ($\Sigma\sbr{GC}$) at the effective radius $R\sbr{e}$, the radius that encloses half of the total GC density. The parameter $n$ describes the shape of the curve, and $b_n$ is a constant that is dependent on $n$. The constant, $b_n$, is calculated using $b_n = 1.9992n - 0.3271$ \citep{Graham05}. The \sersic fit can be used to calculate the total number of GCs~($N\sbr{GC}$) around the galaxy by integrating \equref{sersic} over a projected 2D area to obtain
\begin{equation}
N\sbr{GC} = \Sigma\sbr{e} R\sbr{e}^2 2\pi n\frac{e^{b_n}}{(b_n)^{2n}}\Gamma(2n).
\end{equation}

In addition to a \sersic profile, the galaxies were also fit with a power law of the form
\begin{equation}
\label{eq:powerlaw}
\Sigma\sbr{PL}(R) = \Sigma_0\left(\frac{R}{R_0}\right)^{\gamma}.
\end{equation}
$\Sigma_0$ and $\gamma$ were treated as free parameters. To avoid degeneracy, $R_0$ was set to a fixed value of 100\,kpc. Normally, this model has an additional term that represents a core within which the number density begins to flatten. However, we will be limiting our GC detection aperture with a certain inner radius (discussed in \secref{modelfit}) that is much larger than the core radius \citep{Forbes96}. Consequently the core radius term is omitted in this power law model. The power law model can also be used to calculate the number of GCs belonging to each galaxy. This is done simply by integrating \equref{powerlaw} over a projected 2D area to obtain
\begin{equation}
\label{eq:powerint}
N\sbr{GC}=\int_{R\sbr{min}}^{R\sbr{max}}\!2\pi\!R\,\Sigma\sbr{PL}(R)\,dR
= \frac{2 \pi \Sigma_{0}}{R_0^{\gamma} (\gamma +2)} (R\sbr{max}^{\gamma+2} -
R\sbr{min}^{\gamma+2})
\end{equation}

\subsection{GCS profiles and fits}
\label{sec:modelfit}

In this work, our primary goal is to understand the spatial extent of the GC system, particularly on large scales. Because the hosts are often not smooth early types, we do not attempt to model and subtract the galaxy light in order to identify GCs close to the galaxy centre. Instead we first adopt an inner radius, $R\sbr{in}$, which is taken to be the $K$-band 20$^{\textrm{th}}$ magnitude isophotal radius from the 2MASS All-Sky Extended Source Catalog \citep{Jarrett00}. We then calculate surface density profiles using concentric, logarithmically-scaled annular bins  extending from $R\sbr{in}$  to the edge of the image.  As seen in \figref{density}, $R\sbr{in}$ ranges from $5-15$\,kpc. To avoid contamination from the globular cluster systems of other galaxies, circular masks of radius $0.15~\times~R_{200}$ were placed around each other galaxy and added to the overall mask for the galaxy being studied. The number of GCs in each annulus was divided by the unmasked area of the annulus to obtain the number density, yielding the profile of GC surface number density versus galactocentric distance for each galaxy \citep[as][]{Faifer11, Salinas15}, based on  the distance to each galaxy from \tabref{important}. Surface density profiles for globular clusters are shown in \figref{density}. The number of red and blue GCs were also counted in each annulus separately, using $(g-i)$ = 0.80 to separate the subpopulations (\secref{colourselect}).  Red and blue surface density profiles are shown in \figref{density} as red squares and blue triangles, respectively.

Three different radial profiles were fit to the surface density of GCs as a function of projected radius. These models are (1) a de Vaucouleurs ($n=4$) profile where $R\sbr{e}$ was allowed to vary, and (2) a power law, and (3) de Vaucouleurs profile where $R\sbr{e}$ was kept constant.  A constant background term was included in the fit. After subtracting the background, we use the fraction of GCs within the magnitude limits (as discussed in \secref{candidates}) and divide the raw GC density in each bin by this fraction to obtain surface density profiles that are corrected for background contamination and incompleteness.

In the free $R\sbr{e}$ de Vaucouleurs fits there is considerable degeneracy between the fitted $R\sbr{e}$ and fitted $\Sigma\sbr{e}$, leading to large uncertainty in the total counts. Therefore, we also perform fits fixing $R\sbr{e} = 0.05 R_{200}$.  We compare the corrected GC counts within an annulus $R\sbr{in}$ to $R\sbr{out} = 100$ kpc to the integral of the models over the same annulus. The comparison of these aperture counts is given in  \tabref{counts}. In most cases, the direct, corrected counts agree well with the models within the aperture.  This indicates that our models are reasonable within the range galactocentric radii $R\sbr{in} <  R  < R\sbr{out} = 100$ kpc.

The fit parameters, as well as the $\chi^2$ of the fits are given in \tabref{redevauc} for the free-$R\sbr{e}$ fits, and in Appendix~\ref{sec:parameters} for the other profiles. The $\chi^2$ allows us to compare these three models.
The variable-$R\sbr{e}$ de Vaucouleurs fit and the power law fit both have the same number of free parameters. In most cases, the de Vaucouleurs fit has a slightly lower $\chi^2$, but the difference is usually negligible ($ \lesssim1$) suggesting that our data do not easily distinguish these two profiles. In the case of NGC~942+943, the variable $R\sbr{e}$-de Vaucouleurs fit does offer a significantly better $\chi^2$ and appears to be the better fit. The blue GCs of NGC~942+943 seem to be slightly better represented by the \sersic model. \cite{Faifer11} also fit both a de Vaucouleurs profile and a power law to GC surface density distributions and also found the two models yield very similar quality of fit. However, they found that the inner regions are represented slightly better by the de Vaucouleurs profile. The power law fit has 13 degrees of freedom, while the fixed $R\sbr{e}$-de Vaucouleurs fit has 14. In most cases there is not a significant difference in the quality of the fit of the two models. NGC~883 is slightly better fit by the power law, but this difference is less significant than the difference between the fixed $R\sbr{e}$ and the variable $R\sbr{e}$ de Vaucouleurs fits for this galaxy. Overall, the GC surface density distributions are well represented by all three models.

\subsection{Total Number of Globular Clusters: Comparison to Previous Results}
\label{sec:counts}

\begin{figure*}
\centering
\includegraphics[scale = 1]{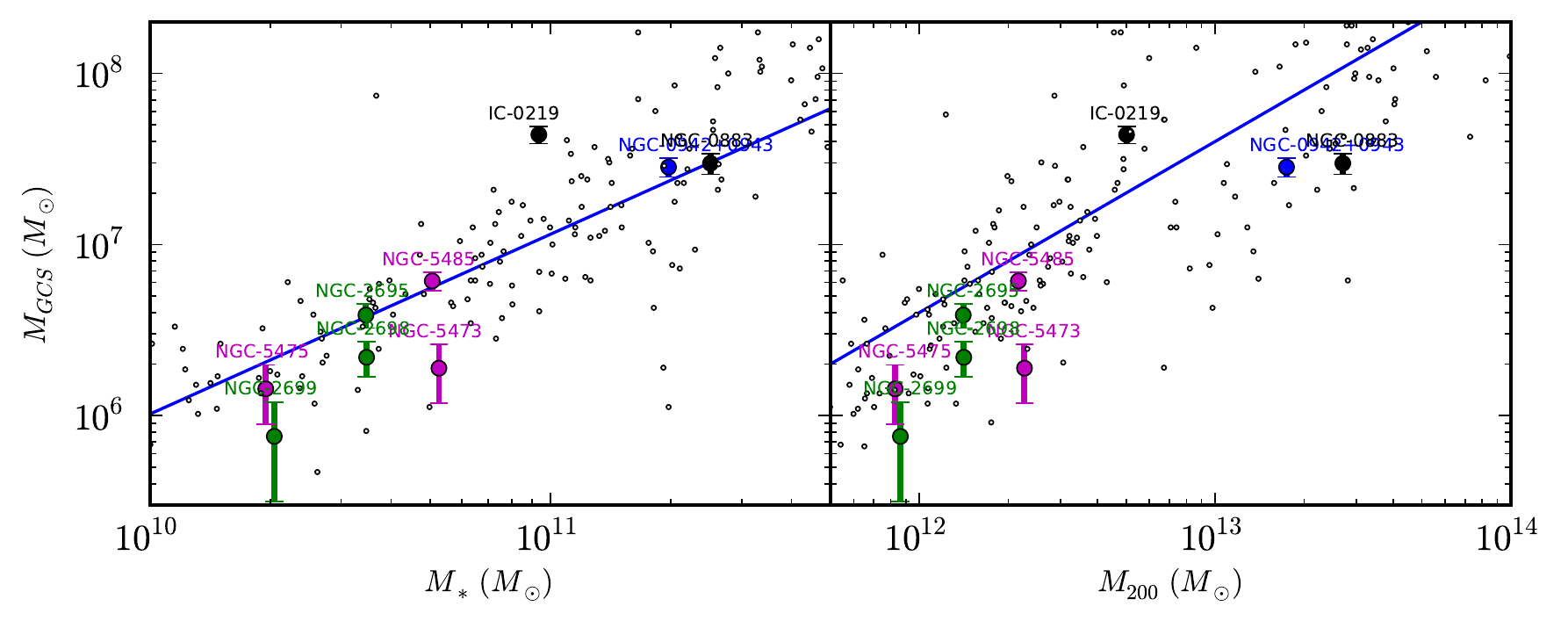}
\caption{The relationship between $M\sbr{GCS}$ and $M_*$ (left hand panel) or $M_{200}$ (right hand panel). $M\sbr{GCS}$ was determined using the extrapolated counts described in \protect\secref{counts}. \protect\cite{Harris13} galaxies are represented by small gray points. Galaxies from this paper are represented as circles labelled according to their name. Galaxies that are members of the same group are given the same colour. A line of slope~$1.04$ is plotted on the left side of \figref{GCSmass} to represent this relationship. In the right panel, the blue line indicates the relation of  \protect\cite{Hudson14}: $\eta = M\sbr{GCS}/M_{200} = 4 \times 10^{-5}$.}
\label{fig:GCSmass}
\end{figure*}

The main purpose of this paper is to compare the spatial sizes of GC systems. However, in order to compare our results with previous work, it is useful to also obtain total GC counts and red GC fractions. To determine the total number of GCs across all galactocentric radii, the fixed-$R\sbr{e}$ de Vaucouleurs fits GCs in the inner region was calculated by evaluating the de Vaucouleurs integral from $R\sbr{min} = 0$\,kpc to $R\sbr{max} = R\sbr{in}$. The number of GCs in the outer region was calculated by integrating the de Vaucouleurs fit from $R\sbr{min} = R_{out}$ to $R\sbr{max} = \infty$ and integrating. The extrapolated inner regions, the aperture counts, and the extrapolated outer regions were added together to create total ``combined'' counts. The counts determined through the various methods in this section are shown in \tabref{counts}, including the  combined counts for red and blue GCs.

As discussed in the introduction, the total mass of the GCS is related to both the stellar mass of the host galaxy and to its halo mass.  The total mass of the GCS system for our galaxies, $M\sbr{GCS}$, was obtained by multiplying the combined counts by the average GC mass of $2.4 \times 10^5 M_{\sun}$ \citep{Durrell14}. In \figref{GCSmass}, $M\sbr{GCS}$ of each GCS is plotted against $M_*$ and $M_{200}$ of its host galaxy and these results are compared to galaxies from the compilation of \cite{Harris13}.

\begin{figure*}
\centering
\includegraphics[scale = 1]{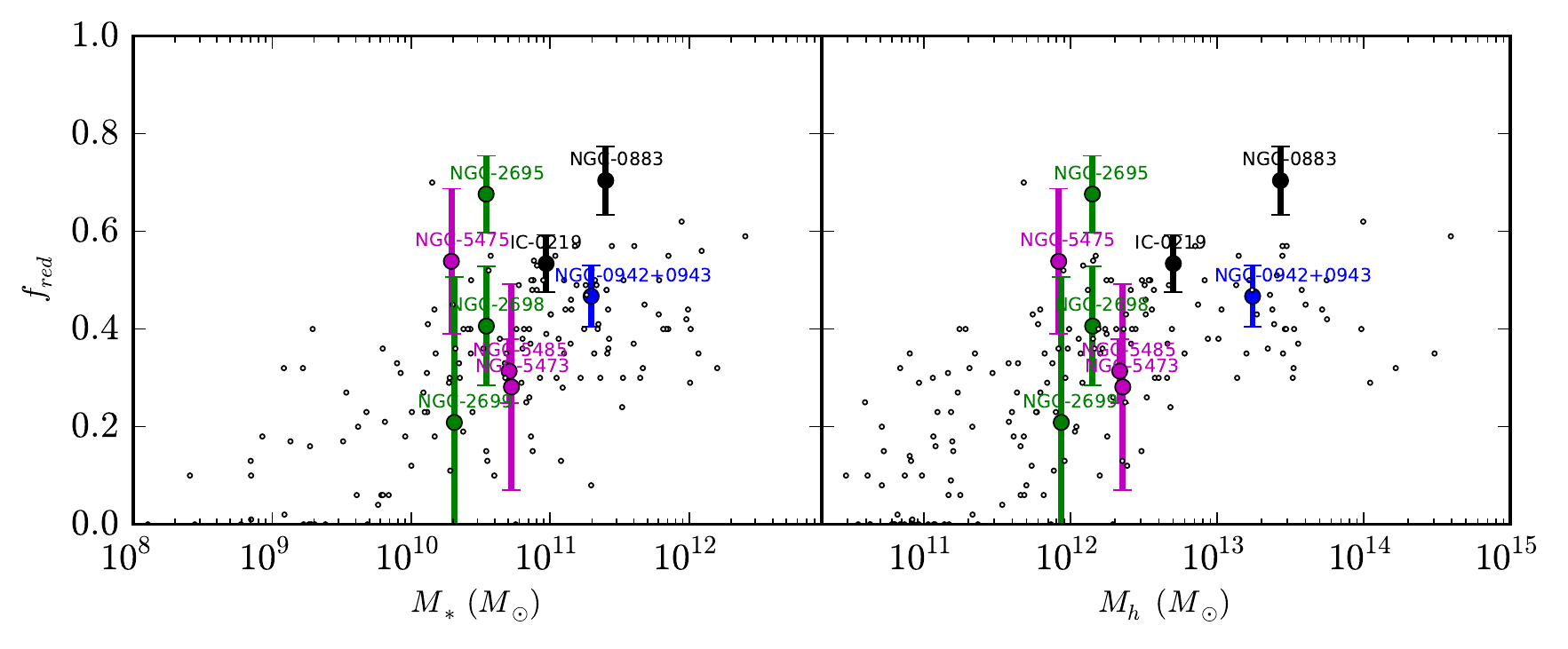}
\caption{The red fraction $f\sbr{red}$ for galaxies in this paper compared with data from \protect\cite{Harris15}. Red fractions were calculated using the de Vaucouleurs fit. $f\sbr{red}$ is plotted against $M_*$ in the left panel and $M_{200}$ in the right panel.}
\label{fig:fred}
\end{figure*}
We also compare the ratio of red GCs to total GCs, or red fraction ($f\sbr{red}$), between our galaxies and the Harris catalogue galaxies. $f\sbr{red}$ was calculated using $N\sbr{GC}$ determined by the combined counts method of de Vaucouleurs extrapolation. In \figref{fred}, $f\sbr{red}$ of each GCS is plotted against $M_*$ and $M_{200}$ of the parent galaxy. The galaxies in this paper generally follow the overall trends from \cite{Harris13}, although a couple (NGC 2695 and NGC 883) are at red end of the distribution for their stellar or halo mass.

\section{Spatial Extent of Globular Clusters}
\label{sec:spatial}

\begin{figure*}
\centering
\includegraphics[scale = 1]{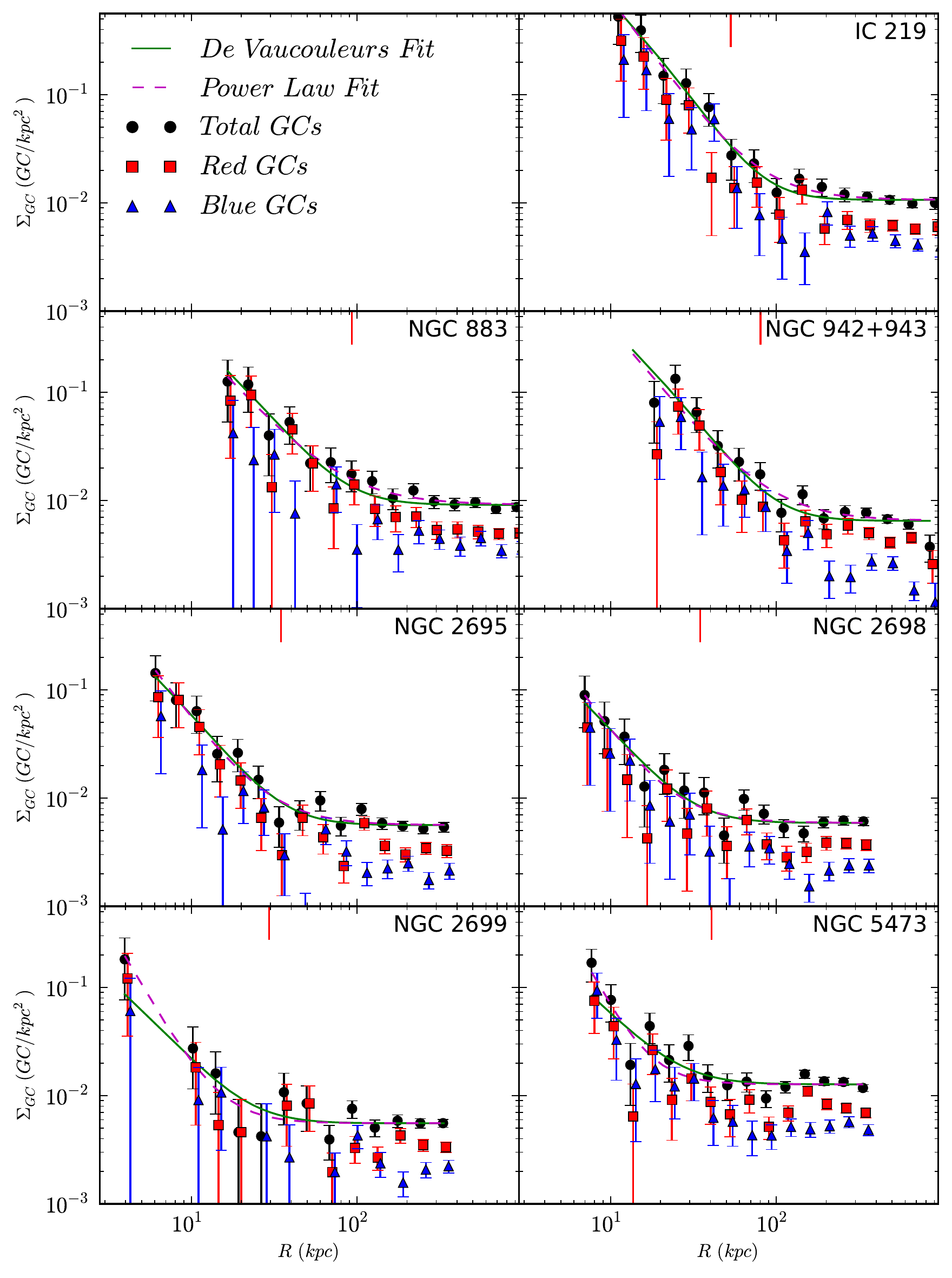}
\end{figure*}

\begin{figure*}
\centering
\includegraphics[scale = 1]{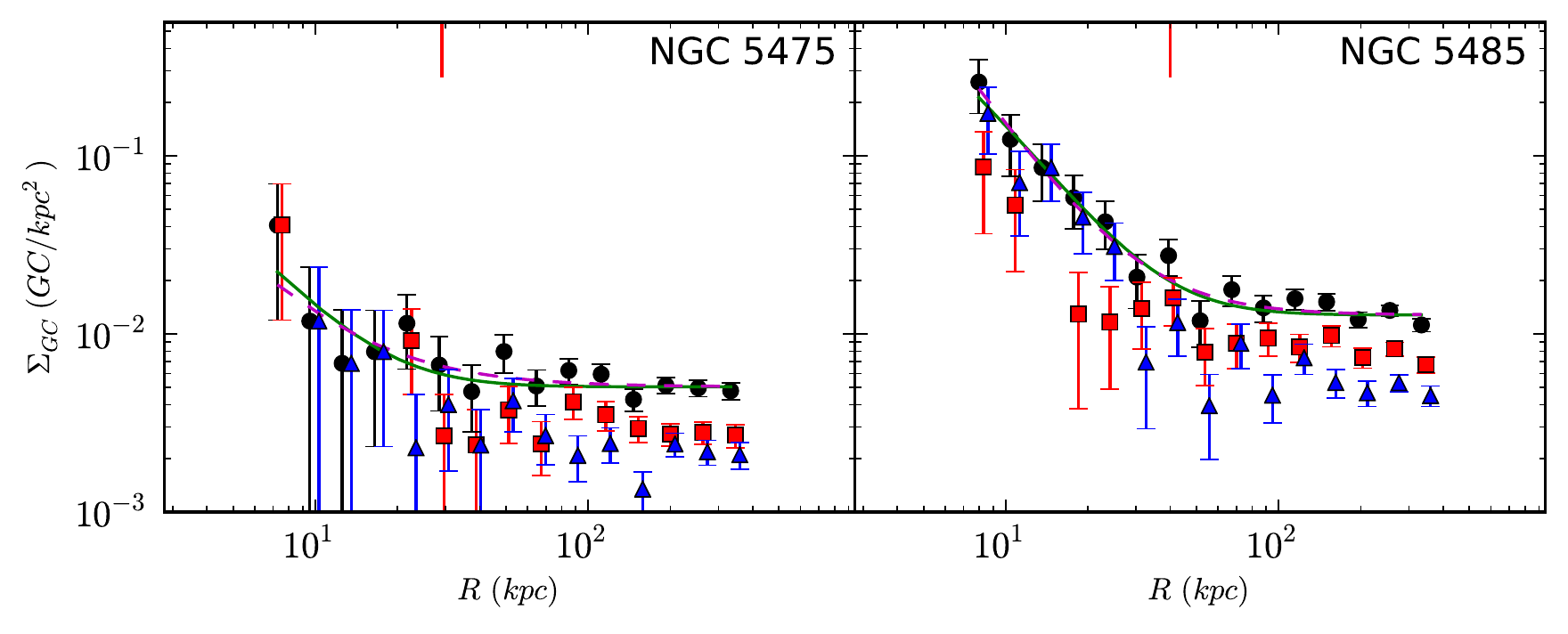}
\caption{Surface density of GCs, $\Sigma\sbr{GC}$, in $\frac{\textrm{GC}}{\textrm{kpc}^2}$, versus galactocentric distance ($R$, in kpc) for all galaxies. Total GC counts are displayed as black circles, red GCs as red triangles, and blue GCs as blue squares. The de Vaucouleurs fit appears a solid green line, while the power law fit appears as a dashed purple line. The red tick mark at the top of each plot corresponds to a galactocentric distance of 3\,$R\sbr{e}$ or $0.15\times R_{200}$.}
\label{fig:density}
\end{figure*}

\begin{table*}
\caption{The fit parameters of free-$R\sbr{e}$ de Vaucouleurs fits for galaxies studied in this paper, as well as the $\chi^2$ of the fit. In each fit there are 13 degrees of freedom.}
\begin{tabular}{lccccccccc}
\hline
Galaxy&Total $\Sigma\sbr{e}$&Total $R\sbr{e}$&Total $\chi^2$&Red $\Sigma\sbr{e}$&Red $R\sbr{e}$&Red $\chi^2$&Blue $\Sigma\sbr{e}$&Blue $R\sbr{e}$&Blue $\chi^2$\\
&$(10^{-2}\frac{GC}{kpc^2})$&(kpc)&&$(10^{-2}\frac{GC}{kpc^2})$&(kpc)&&$(10^{-2}\frac{GC}{kpc^2})$&(kpc)&\\
\hline
IC 219&11.84$\pm$8.17&25.89$\pm$8.83&5.99&16.11$\pm$19.72&15.70$\pm$8.72&6.75&8.71$\pm$11.41&19.46$\pm$12.15&8.67\\
NGC 883&0.28$\pm$0.17&177.49$\pm$71.17&4.39&0.10$\pm$0.09&259.53$\pm$167.55&6.96&0.12$\pm$0.21&125.87$\pm$136.78&6.21\\
NGC 942+943&1.04$\pm$1.07&77.53$\pm$44.90&13.81&0.36$\pm$0.53&89.71$\pm$77.66&11.58&0.30$\pm$0.34&106.82$\pm$69.42&10.80\\
NGC 2695&5.05$\pm$6.16&10.32$\pm$6.08&11.06&14.39$\pm$26.82&5.30$\pm$4.23&15.01&26.82$\pm$117.77&2.90$\pm$4.82&16.92\\
NGC 2698&3.50$\pm$6.82&10.35$\pm$9.86&9.85&15.25$\pm$58.53&3.52$\pm$5.31&7.53&9.81$\pm$26.13&4.87$\pm$5.47&9.03\\
NGC 2699&77.93$\pm$162.14&2.01$\pm$1.62&6.34&62.15$\pm$186.21&1.83$\pm$2.08&7.85&0.15$\pm$0.34&32.85$\pm$47.00&6.31\\
NGC 5473&539.84$\pm$2.55e3&1.55$\pm$2.37&18.99&1.07e6$\pm$1.29e7&0.18$\pm$0.43&18.82&4.93$\pm$10.65&7.86$\pm$7.79&5.85\\
NGC 5475&0.01$\pm$0.01&423.42$\pm$567.03&6.04&28.06$\pm$87.50&2.93$\pm$3.52&5.68&0.05$\pm$0.15&50.05$\pm$89.11&7.76\\
NGC 5485&8.52$\pm$9.89&12.85$\pm$7.27&11.29&0.01$\pm$0.02&449.30$\pm$501.26&6.80&24.67$\pm$36.41&6.52$\pm$4.09&12.32\\
\end{tabular}
\label{tab:redevauc}
\end{table*}

In this section, we investigate the radial distribution of the GCS as a function of their host galaxy or halo properties.  We will combine GCS sizes from the CFHTLS galaxies studied in this paper with data from the literature.

\subsection{Fits to the CFHTLS galaxy sample}

\begin{figure}
\centering
\includegraphics[width=0.45\textwidth]{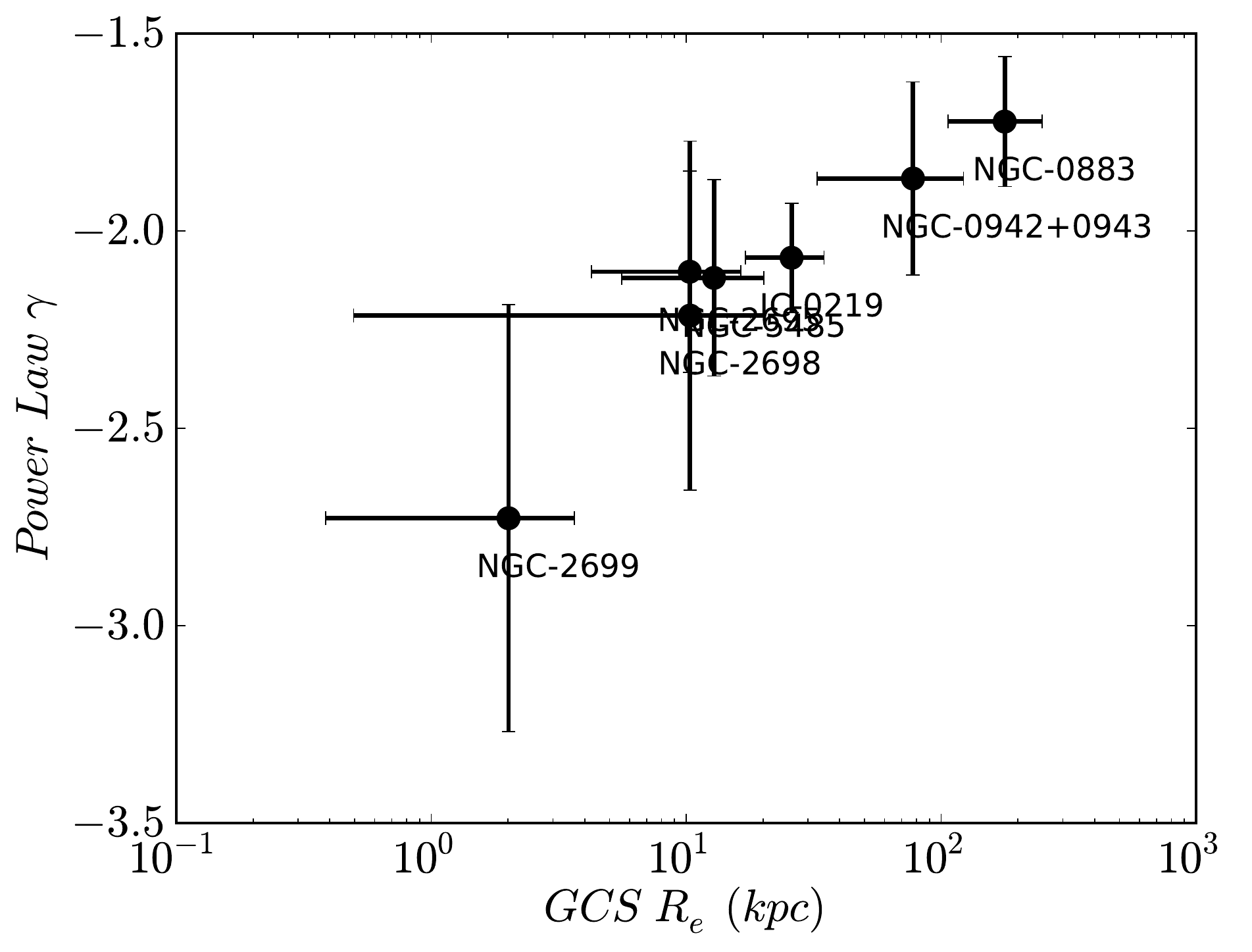}
\caption{A plot of de Vaucouleurs profile fit $R\sbr{e}$ vs power law fit $\gamma$.}
\label{fig:gammaRe}
\end{figure}

We have fit a free $R\sbr{e}$ to the GCS of galaxies in our sample, assuming de Vaucouleurs profile. The results are tabulated in \tabref{redevauc} for all GCs, including for red and blue GCs separately. The GCSs of two of our galaxies, NGC~5473 and NGC~5475, do not have a well-defined $R\sbr{e, GCS}$. Therefore, we omit these from subsequent discussion in this section.  We also fit the power law model: results are in \tabref{fixparam}. The power-law and de Vaucouleurs fits are compared in \figref{gammaRe}.  As expected, more extended galaxies with a larger $R\sbr{e}$ have a less negative $\gamma$ and the two parameters are correlated. 

\begin{figure}
\includegraphics[width=0.45\textwidth]{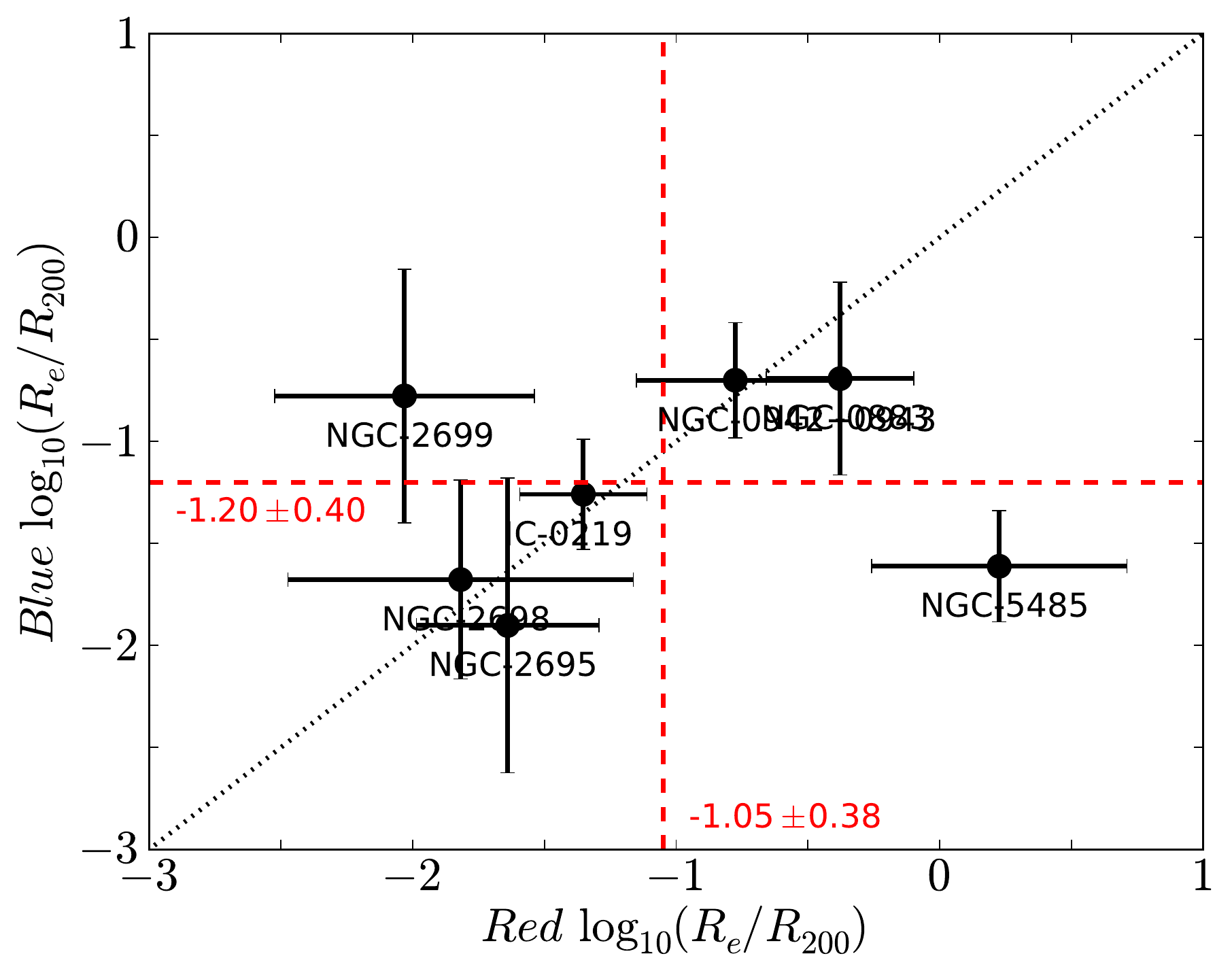}
\includegraphics[width=0.45\textwidth]{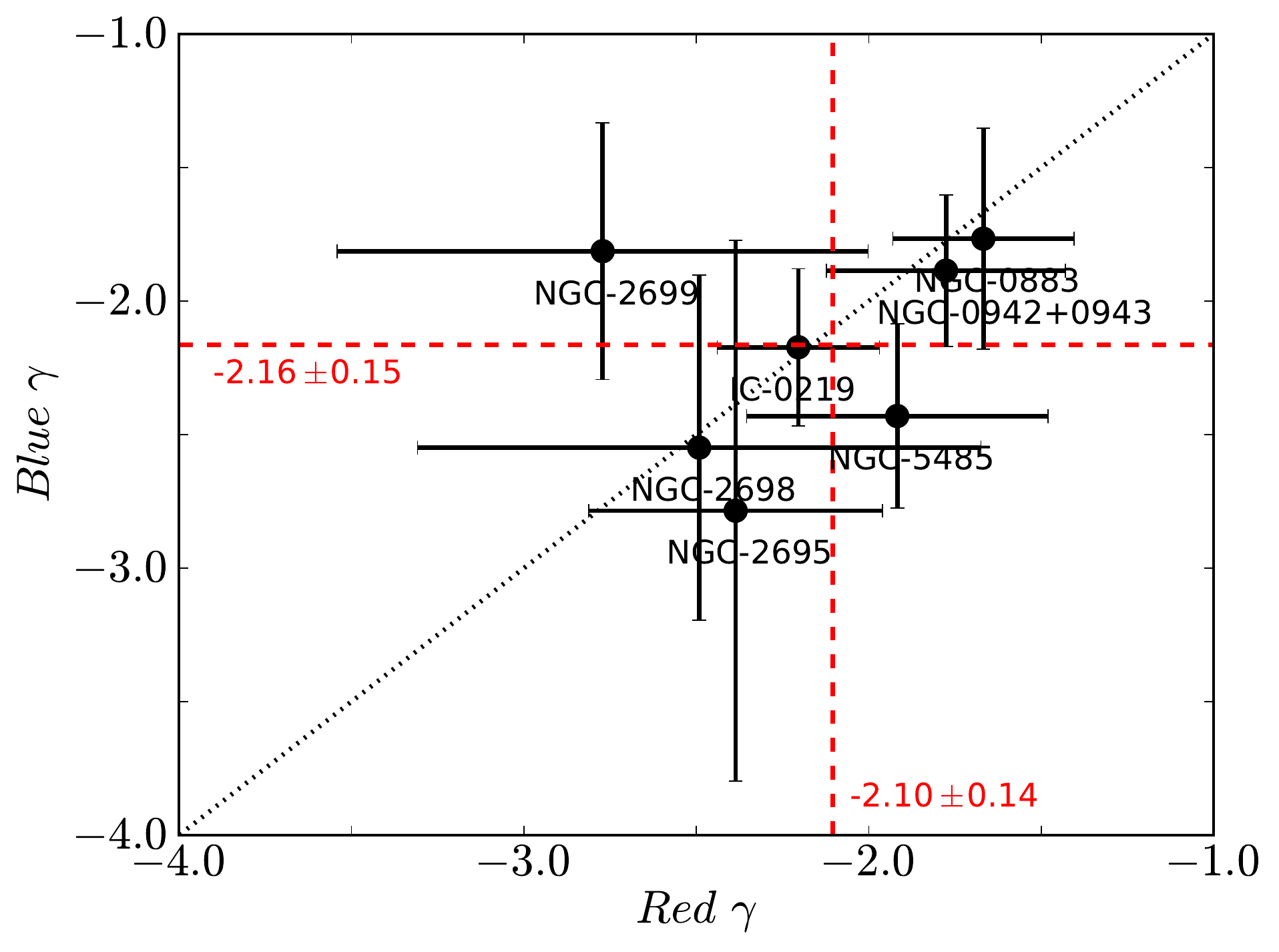}
\caption{Comparison of red GC and blue GC distributions. In each plot, the dotted black line represents the line $y = x$. The vertical red dashed line is the average for red GCs and the horizontal red dashed line is the average for blue GCs. Top: Comparison of \ratio\,\! for red and blue GCs. Bottom: Comparison of $\gamma$ from the power law fit for red and blue GCs.}
\label{fig:rbcompare}
\end{figure}

It is also interesting to study spatial profiles of the the red and blue GCS populations. In most previous studies, the blue GCS distribution is more extended that the red GCS distribution \citep{Rhode04, BasFaiFor06, Faifer11, Kartha14, ChoBlaChi16}. A comparison of the red and blue GC profiles for the galaxies studied in this paper is shown in \figref{rbcompare}. The upper panel compares \ratio\,\! ratios of both populations. The lower panel of the figure compares $\gamma$ from the power law fit for both populations.  If the blue population is more extended, one would expect \ratio\,\! to be higher for blue GCs and $\gamma$ to have a larger negative value for red GCs. This is the sense of the observed trend, but the uncertainties on the parameters of the red and blue parameters are large, and so within the errors, they are also consistent with being equal.

\subsection{Data for other GCS}

Our CFHTLS sample contains only 7 galaxies with usable GCS $R\sbr{e}$ measurements. Here we describe additional measurements of GCS sizes that allow us to extend the dynamical range of the sample. Specifically, we describe our analysis of the Milky Way, M31 and M87/Virgo, as well as additional galaxies from the literature.

\begin{figure*}
\centering
\includegraphics[scale = 1]{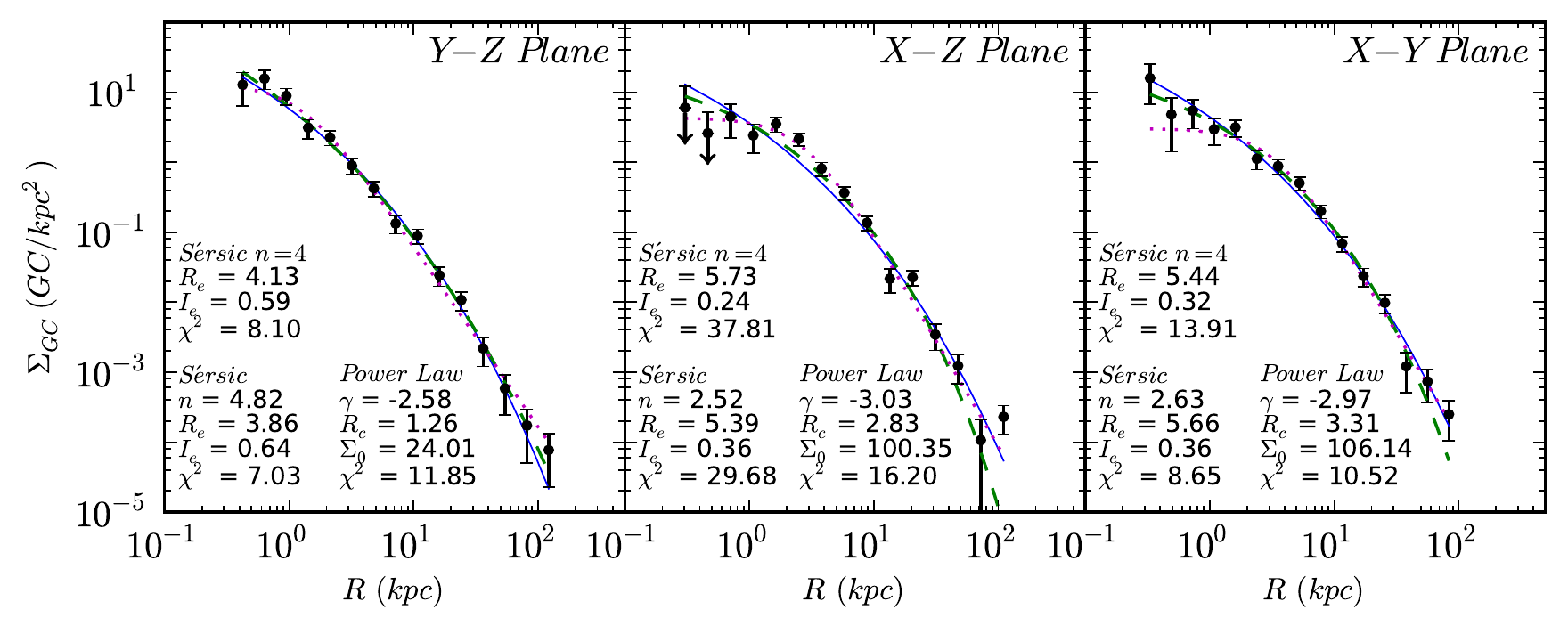}
\caption{The number density of Milky Way GCs \protect\citep[from][]{Harris96} as would be observed in projection onto the three Cartesian planes in Galactic coordinates. A de Vaucouleurs fit ($n = 4$) is shown by a solid blue curve. A \sersic (free-$n$) fit is represented by a dashed green line. A power law fit shown as a dotted purple line. The error bars on the first two points in the middle plot actually extend to the edge of the plot, but are represented by arrows to avoid overlapping with the fit parameters.}
\label{fig:mway}
\end{figure*}

\subsubsection{The Milky Way GCS in projection}
\label{sec:milkyway}
To obtain the GCS effective radius for the Milky Way (MW), we use  the catalogue of MW GCs has been compiled by \cite{Harris96} and updated in \cite{Harris10}. We retrieve the Galactic Cartesian $X$, $Y$, and $Z$ positions of each GC relative to the Galactic centre, and use these to create 2D projections of the GCS in the $XY$, $XZ$, and $YZ$ planes. The GCs were placed into 15 concentric annular logarithmically-space bins by projected galactocentric distance.  The three projections of surface density are shown in \figref{mway}.
The fit to a de Vaucouleurs profile is shown in \figref{mway} as a solid blue line.  The $Y$$Z$ plane is the least likely of the three projections to be affected by extinction towards the Galactic centre, so we adopt this fit as the preferred value: $R\sbr{e} = 4.1$ kpc. We estimate an uncertainty of $0.5$ kpc.
This is similar to the value 4.4 kpc obtained by \cite{Battistini93}. 
A  \sersic (free $n$) profile was also fit and is shown in \figref{mway} as the dashed green line. 
The average $n$ was $3.3$ but in the $YZ$ plane $n = 4.82$. 
The $\chi^{2}$ of this fit is not significantly better than $n = 4$.  
We conclude there is no strong evidence for deviations from $n=4$ in the MW GCS. 
With free $n$, the $R\sbr{e}$ for the $YZ$ projection is  $3.86$\,kpc.
Finally, a cored power law model, with the functional form
\begin{equation}
\Sigma\sbr{PL}(R) = \Sigma_0\left(R_c^2+R^2\right)^{\frac{\gamma}{2}}.
\label{eq:powercore}
\end{equation}
was fit to the MW GCS data. In the above equation, $R_c$ represents the core radius. The core radius, $R_c$, was found to be $1.3$\,kpc and the power-law $\gamma$ was $-2.6$, again for the $YZ$ plane. This is a slightly poorer fit than the de Vaucouleurs profile. 

Finally, the stellar mass of the Milky Way ($5\pm1 \times 10^{10} M_{\odot}$) was obtained from \cite{BlaGer16}.

\makeatletter{}\begin{table*}
\caption{Globular cluster counts determined using different methods. Within the aperture, the counts are calculated by summing annuli over the specified range, using the de Vaucouleurs fit, and by using the power law fit. The combined extrapolated fit is presented here for total, red, and blue GCs. The red fraction for combined counts is also presented here.}
\begin{tabular}{lccccccc}
\hline
&\multicolumn{3}{c|}{Aperture}&\multicolumn{4}{c}{Combined}\\
&\multicolumn{3}{c|}{$R\sbr{in}~<~R~<~100$\,kpc}&\multicolumn{4}{c}{$0~<~R~<~\infty$}\\
Galaxy&Aperture&de Vaucouleurs Fit&\multicolumn{1}{c|}{Power Law Fit}&Red&Blue&$f_{red}$&Total\\
\hline
IC 219&1046$\pm$192&1032$\pm$120&1041$\pm$115&937$\pm$151&817$\pm$141&0.53$\pm$0.06&1829$\pm$205\\
NGC 883&720$\pm$157&532$\pm$87&553$\pm$75&806$\pm$138&338$\pm$97&0.70$\pm$0.07&1242$\pm$169\\
NGC 942+943&647$\pm$124&641$\pm$130&624$\pm$129&532$\pm$103&607$\pm$96&0.47$\pm$0.06&1181$\pm$147\\
NGC 2695&119$\pm$25&79$\pm$13&86$\pm$18&110$\pm$21&53$\pm$16&0.68$\pm$0.08&161$\pm$25\\
NGC 2698&59$\pm$20&52$\pm$11&55$\pm$18&36$\pm$15&53$\pm$14&0.41$\pm$0.12&91$\pm$21\\
NGC 2699&21$\pm$18&27$\pm$10&26$\pm$11&7$\pm$13&29$\pm$12&0.21$\pm$0.30&31$\pm$18\\
NGC 5473&38$\pm$27&66$\pm$21&49$\pm$22&22$\pm$21&56$\pm$18&0.28$\pm$0.21&78$\pm$29\\
NGC 5475&50$\pm$22&12$\pm$5&22$\pm$11&41$\pm$17&35$\pm$15&0.54$\pm$0.15&59$\pm$22\\
NGC 5485&130$\pm$26&193$\pm$29&208$\pm$38&77$\pm$21&168$\pm$23&0.31$\pm$0.07&255$\pm$31\\
\end{tabular}
\label{tab:counts}
\end{table*}

\subsubsection{M31 GCS}

\begin{figure}
\centering
\includegraphics[width=\columnwidth]{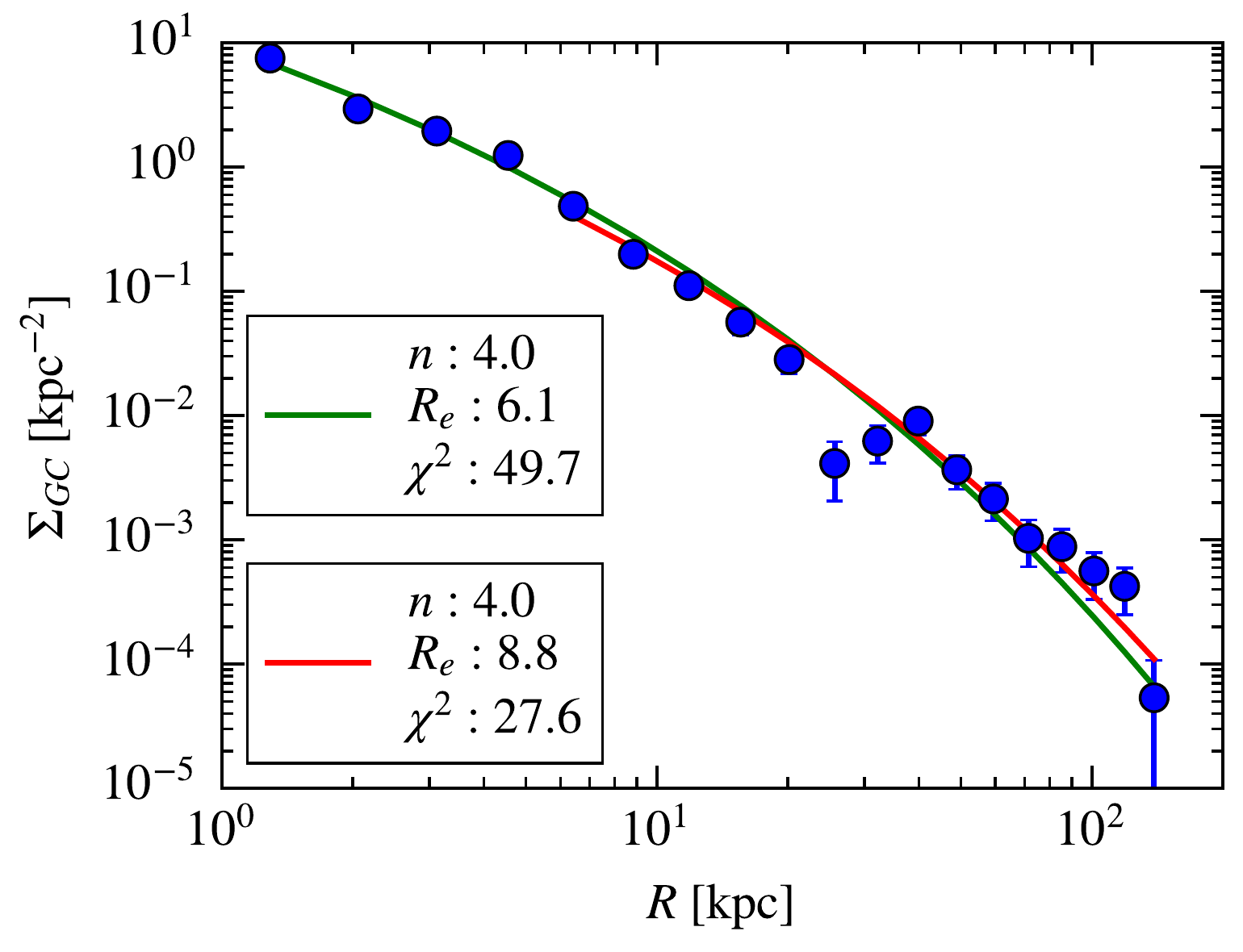}
\caption{GCS surface density as a function of radius for M31.  Fits for two different radial ranges are shown.}
\label{fig:M31}
\end{figure}

We have also produced an updated fit to the M31 GCS effective radius. 
Specifically, we use the GC compilation from \cite{CalRom16}.
The GC surface density profile is shown in \figref{M31}.  \cite{HuxFerTan11} noted that M31 was not well fit by a single profile. Specifically they studied power laws with three components and Sersic profiles with two components. Our goal here to treat M31 in a similar way to the other galaxies in our sample. Hence we restrict the fit to counts at radii larger than the isophotal radius at 20th mag per square arcsecond in the $K$-band, 
which at the adopted distance of 780 kpc corresponds to 6.14 kpc \citep{JarCheCut03}.

Neither power law nor S\'{e}rsic profiles are very good fits to the entire range: the data appear to oscillate around any smooth function. It is not clear whether this is a result of incompleteness in the GC catalogues at some radii. The best fit to the outer regions has $n \sim 4$  and $R\sbr{e} = 8.8\pm0.3$ kpc.  This fit clearly overpredicts the counts in the inner regions. (Had we used the whole range, we would have obtained $R\sbr{e} = 6.1\pm0.8$ kpc.)
The adopted stellar mass of M31 is $1.0\pm0.2\ten{11}$, from \cite{Sick15}.

\subsubsection{M87/Virgo}

To extend the dynamic range of the GCS sample to include dark matter haloes with the highest masses, we include the GCS of M87. Recall that, as discussed in \secref{galaxies}, for central cluster galaxies such as M87, the ``host halo'' is the entire galaxy cluster.  Virgo is a galaxy cluster for which there are GC data spanning the entire cluster. Specifically, \cite{Lee10} studied the GCS of Virgo, and fit its GC density profile with a broken power law.  We have used their parametric fit to estimate the projected radius which encloses half of the Virgo GCs within 5 degrees of M87. We obtain a GCS effective radius of $84\pm56$ kpc, where the uncertainty is estimated via the scatter in the power law parameters.  We take the stellar mass of M87 to be  $5.5^{+1.5}_{-2.0} \times 10^{11} M\sun$ \citep{AgnEvaRom14}.

\subsubsection{Other GCS}
\makeatletter{}\begin{table*}
\caption{GCS and galaxy parameters, including GCS $R\sbr{e}$, galaxy $R\sbr{e}$, stellar mass, halo mass, for each galaxy. Galaxies from this paper are listed first, followed by galaxies from the literature. All of the galaxies in this table appear in \figref{RevsM}. For some of the literature galaxies, the value of $R\sbr{e}$ of the GCS was taken directly from the source. In some cases, density data was provided and was fit by a de Vaucouleurs model to determine $R\sbr{e}$ in the same manner as the galaxies in this paper. These galaxies are marked with an asterisk.}\begin{tabular}{lccccccl}
\hline
Galaxy&GCS $R\sbr{e}$&Galaxy $R\sbr{e}$&$M_*$&$M_{200}$&$R_{200}$&Distance&Source of GCS\\
&(kpc)&(kpc)&(M$_{\odot}$)&(M$_{\odot}$)&(kpc)&(Mpc)&\\
\hline
IC 219&25.89$\pm$8.83&1.88&9.353e+10&5.020e+12&353&72.7&--\\
NGC 883&177.49$\pm$71.17&3.83&2.509e+11&2.704e+13&619&72.7&--\\
NGC 942+943&77.53$\pm$44.90&--&1.972e+11&1.744e+13&535&67.0&--\\
NGC 2695&10.32$\pm$6.08&1.15&3.461e+10&1.407e+12&231&26.5&--\\
NGC 2698&10.35$\pm$9.86&0.88&3.475e+10&1.413e+12&231&26.5&--\\
NGC 2699&2.01$\pm$1.62&0.79&2.044e+10&8.624e+11&196&26.5&--\\
NGC 5473&1.55$\pm$2.37&1.58&5.267e+10&2.267e+12&271&26.2&--\\
NGC 5475&423.42$\pm$567.03&1.31&1.946e+10&8.281e+11&193&26.2&--\\
NGC 5485&12.85$\pm$7.27&2.00&5.067e+10&2.162e+12&266&26.2&--\\
MW & 4.1$\pm$0.5&--&5.0e+10&2.12e+12&265&--&--\\
M31& 8.8$\pm$0.3&--&1.030e+11&5.827e+12&371&0.8&--\\
M87& 87$\pm$56&--&5.5e11&1.2040e14&1020&16&--\\
NGC 720&13.70$\pm$2.20&4.60$\pm$0.90&1.106e+11&6.525e+12&385&22.7&\cite{Kartha14}\\
NGC 1023&3.30$\pm$0.90&2.57$\pm$0.50&6.383e+10&2.904e+12&294&10.6&\cite{Kartha14}\\
NGC 1055*&5.54$\pm$4.95&5.35$\pm$1.41&5.235e+10&2.250e+12&270&16.4&\cite{Young12}\\
NGC 1407&25.50$\pm$1.40&8.06$\pm$1.60&1.856e+11&1.566e+13&516&22.3&\cite{Kartha14}\\
NGC 2683*&1.04$\pm$0.49&--&4.606e+10&1.929e+12&256&9.9&\cite{RhoZepKun07}\\
NGC 2768&10.60$\pm$1.80&6.66$\pm$1.30&1.018e+11&5.721e+12&368&19.1&\cite{Kartha14}\\
NGC 3384*&7.32$\pm$4.75&--&4.081e+10&1.679e+12&245&10.9&\cite{Hargis12}\\
NGC 3556*&1.76$\pm$1.01&--&3.110e+10&1.263e+12&222&11.4&\cite{RhoZepKun07}\\
NGC 3607&14.20$\pm$2.00&4.20$\pm$1.00&1.038e+11&5.902e+12&372&19.5&\cite{Kartha16}\\
NGC 3608&9.10$\pm$1.00&3.20$\pm$0.70&5.609e+10&2.453e+12&278&23.8&\cite{Kartha16}\\
NGC 4157*&19.45$\pm$15.09&--&6.474e+10&2.960e+12&296&18.6&\cite{RhoZepKun07}\\
NGC 4278&11.30$\pm$1.50&2.39$\pm$0.50&5.515e+10&2.401e+12&276&15.5&\cite{Kartha14}\\
NGC 4365&41.30$\pm$8.10&5.92$\pm$1.20&1.698e+11&1.339e+13&489&21.1&\cite{Kartha14}\\
NGC 4406&28.20$\pm$1.00&7.60$\pm$0.50&1.577e+11&1.178e+13&469&15.9&\cite{Kartha16}\\
NGC 4472&58.40$\pm$8.00&7.90$\pm$0.80&2.938e+11&3.624e+13&682&15.6&\cite{Kartha16}\\
NGC 4594&16.80$\pm$1.00&3.20$\pm$0.70&2.049e+11&1.868e+13&547&10.6&\cite{Kartha16}\\
NGC 4754*&8.84$\pm$3.52&--&4.793e+10&2.021e+12&260&16.1&\cite{Hargis12}\\
NGC 4762*&4.74$\pm$1.14&--&4.713e+10&1.981e+12&259&15.3&\cite{Hargis12}\\
NGC 5813&36.60$\pm$3.00&8.80$\pm$0.80&1.532e+11&1.120e+13&461&28.3&\cite{Kartha16}\\
NGC 5866*&8.47$\pm$1.84&--&4.103e+10&1.690e+12&245&11.7&\cite{Hargis12}\\
NGC 7331*&3.55$\pm$7.82&--&1.317e+11&8.667e+12&423&14.1&\cite{RhoZepKun07}\\
NGC 7332*&1.38$\pm$0.32&1.93$\pm$0.53&1.778e+10&7.701e+11&189&13.2&\cite{Young12}\\
NGC 7339*&0.66$\pm$0.94&2.44$\pm$0.64&3.170e+10&1.287e+12&224&22.7&\cite{Young12}\\
\end{tabular}
\label{tab:radii}
\end{table*}

We supplement the above with additional data from the literature: 11 galaxies from Rhode and collaborators \citep{RhoZepKun07, Hargis12, Young12} and 12 galaxies from Kartha and collaborators \citep{Kartha14, Kartha16}.  For the data from Rhode and collaborators, we refit their GC radial density profile data to obtain $R\sbr{e}$ in a consistent way. 

For all of these galaxies, $K$-band magnitudes, $B-V$~colours, and distances were compiled from NED in order to calculate stellar masses using the methods outlined in \secref{galaxies}. $M_{200}$ and $R_{200}$ of all the above galaxies were obtained using the method introduced in \secref{galaxies}.

\subsection{Results}

In this section, we focus on the de Vaucouleurs fits and compare the $R\sbr{e}$ of the GCS with properties of the host galaxy, such as the $R\sbr{e}$ of its light,  or the halo of which it is a central galaxy. \tabref{radii} summarizes the properties of all the galaxies and their GCS used in this section.

\begin{figure*}
\centering
\includegraphics[scale = 1]{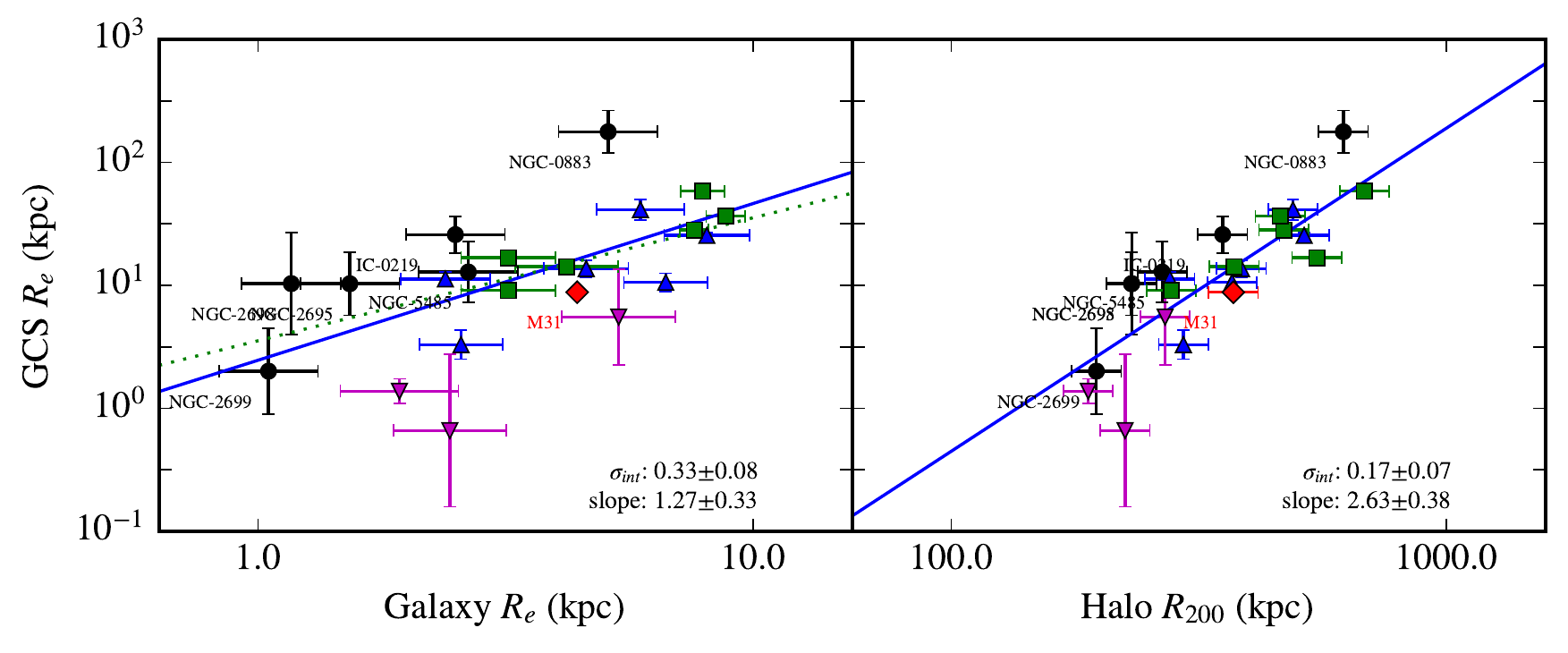}
\caption{GCS $R\sbr{e}$ versus galaxy light $R\sbr{e}$ (left panel) and halo $R_{200}$ (right panel). Galaxies from this paper are black. Galaxies from \protect\cite{Kartha14} are blue triangles. Galaxies from \protect\cite{Kartha16} are green squares. Galaxies from \protect\cite{Young12} are purple inverted triangles. M31 is a red diamond. The line of best fit appears as a solid blue line. In the left panel, the dotted line shows the best-fitting line of slope 1.}
\label{fig:gcsgalRE}
\end{figure*}

\begin{figure*}
\centering
\includegraphics[scale = 1]{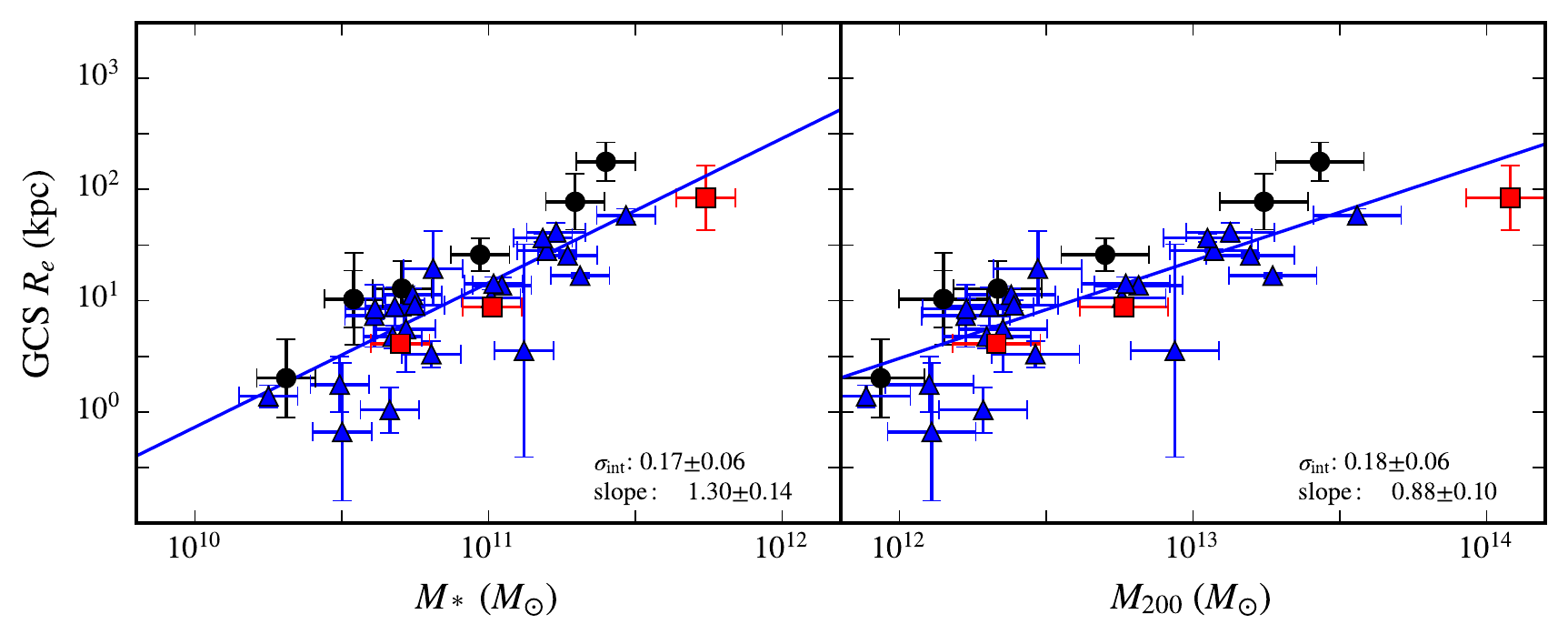}
\caption{GCS $R\sbr{e}$  as a function of galaxy $M_*$ (left) and halo $M_{200}$ (right). The Milky Way, M31 and M87 appear in red, from left to right. Other galaxies from the literature
appear as blue triangles. The solid line is the line of best fit.}
\label{fig:RevsM}
\end{figure*}

\figref{gcsgalRE} (left panel) shows $R\sbr{e}$ of the GCS is plotted against $R\sbr{e}$ of the light for group galaxies studied in this paper, and additional galaxies as described above. The effective radii of CFHLTS galaxies are measured in the $K$-band, we multiply these by a factor 1.33 for consistency with optically-determined effective radii \citep{KoIm05}. We perform a fit allowing for scatter in both the independent and dependent variables. For galaxies with no quoted errors in their effective radii, we assume an error of 0.1 dex.
The line of best fit appears as a dotted black line and is given by
\[
\log_{10} R\sbr{e, GCS} = 1.27 \pm 0.33  \log_{10} R\sbr{e, gal} + 0.39\pm0.22
\]
 The slope is not significantly different from unity, so if we fix it we find that the data are consistent with a constant ratio between the GCS $R\sbr{e}$ and the galaxy light $R\sbr{e}$. Specifically, $R\sbr{e,GCS} ] \sim 3.5 R\sbr{e, light}$. This ratio is lower than the value $6.5 \pm 1.3$ found by \cite{Kartha16} for twelve galaxies, which are a subset of our sample.

It is also interesting to consider the scatter in the relationship.   We measure this by adding a fractional intrinsic scatter, $\sigma\sbr{int}$, to the observational errors in quadrature and fit for this intrinsic scatter via maximum likelihood. From these fits, we find  $\sigma\sbr{int} = 0.33\pm0.08$ dex.

Galaxies are embedded within a dark matter halo and so it is interesting to compare the size of the GCS with that of the halo. The plot on the right side of \figref{gcsgalRE} is similar to the plot on the left, but here GCS $R\sbr{e}$ is plotted against the $R_{200}$ of the dark matter halo instead of the effective radius of the galaxy light. We assume that there is an uncertainty of 0.15 dex in the estimated $M_{200}$ at fixed $M_{*}$, which becomes $0.05$ in $R_{200}$. In the $R_{200}$ plot, the slope is much steeper so that the GCS is not a fixed fraction of the virial radius. Specifically, we find
\[
\log_{10} R\sbr{e, GCS} = 2.63 \pm 0.38  \log_{10} R\sbr{200} - 5.6\pm0.97
\]
More importantly, this comparison has less intrinsic scatter: $\sigma\sbr{int} = 0.17\pm0.07$ dex. This tighter relationship suggests that the $R\sbr{e}$ of the GCS is more closely related to $R_{200}$, and therefore $M_{200}$, than it is to the $R\sbr{e}$ of the host galaxy light.

However, since $M_{200}$ and $R_{200}$ are based on the stellar mass, it is also interesting to see which property correlates best with $R\sbr{e, GCS}$. We have calculated $M_*$ for each galaxy using its $B-V$ colour \citep{Bell03}. The scatter in this relationship between stellar mass at fixed colour is 0.1-0.2 dex, and so we conservatively adopt a scatter in $M_{*}$ of 0.1 dex. Moreover, $M_*$ is used to calculate $M_{200}$ in the manner described by \cite{Hudson15}, who adopt a scatter of 0.15 dex in this relationship as we do here.
\figref{RevsM} shows these comparisons for a larger sample of galaxies. For the relation between $R\sbr{e, GCS}$ and $M_{*}$ we find:
\[
\log_{10} R\sbr{e, GCS} = 1.30 \pm 0.14 \log_{10} (M_{*}/(10^{11} M\sun) + 1.17\pm0.05
\]
whereas for the correlation with $M_{200}$ we obtain
\[
\log_{10} R\sbr{e, GCS} = 0.88 \pm 0.10 \log_{10} (M_{200}/(10^{13} M\sun) + 1.35\pm0.06 .
\]
For both comparisons, the intrinsic scatter is estimated to be $0.17\pm0.06$ dex.  Without direct DM halo measurements, we cannot determine whether $R\sbr{e, GCS}$ correlates better with $M_{*}$ or $M_{200}$.

For the results shown above, our fits were based on de Vaucouleurs ($n = 4$) profiles. However, some literature values are based on free $n$ \sersic fits. If we fit our GCSs with a free $n$, the half-light radii are too noisy to be useful (see discussion in \secref{modelfit}). We have tested the effect of the adopted $n$ on the scaling relations shown in \figref{RevsM}. 
However, the data shown there include some literature galaxies for which $n$ is fixed. Using only the 6 measurements of GCS effective radii from CFHTLS,  we find  that varying $n$ from 1 to 4 causes the slope of the $\log R\sbr{e}$--$\log M_{*}$ relation to change from 0.97 to 1.58 and the slope of the $\log R\sbr{e}$ --$\log M_{200}$ relation varies from 0.70 to 1.15. 
Note also that the CFHTLS effective radii appear to be systematically larger than those from the literature. This may be due to our fitting method which fits only the outer region of the GCS. For example, for the case of M31, we did indeed find a larger $R\sbr{e}$ when the outer regions of the GCS. Alternatively, the differences may be due to differences in the galaxy morphologies of different subsamples. Our CFHTLS sample is mostly early types whereas other samples, notably \cite{Young12} contain relatively more spirals. A larger, more homogeneous sample is needed to understand the systematics in fitting methods.

\begin{figure*}
\centering
\includegraphics[scale = 1]{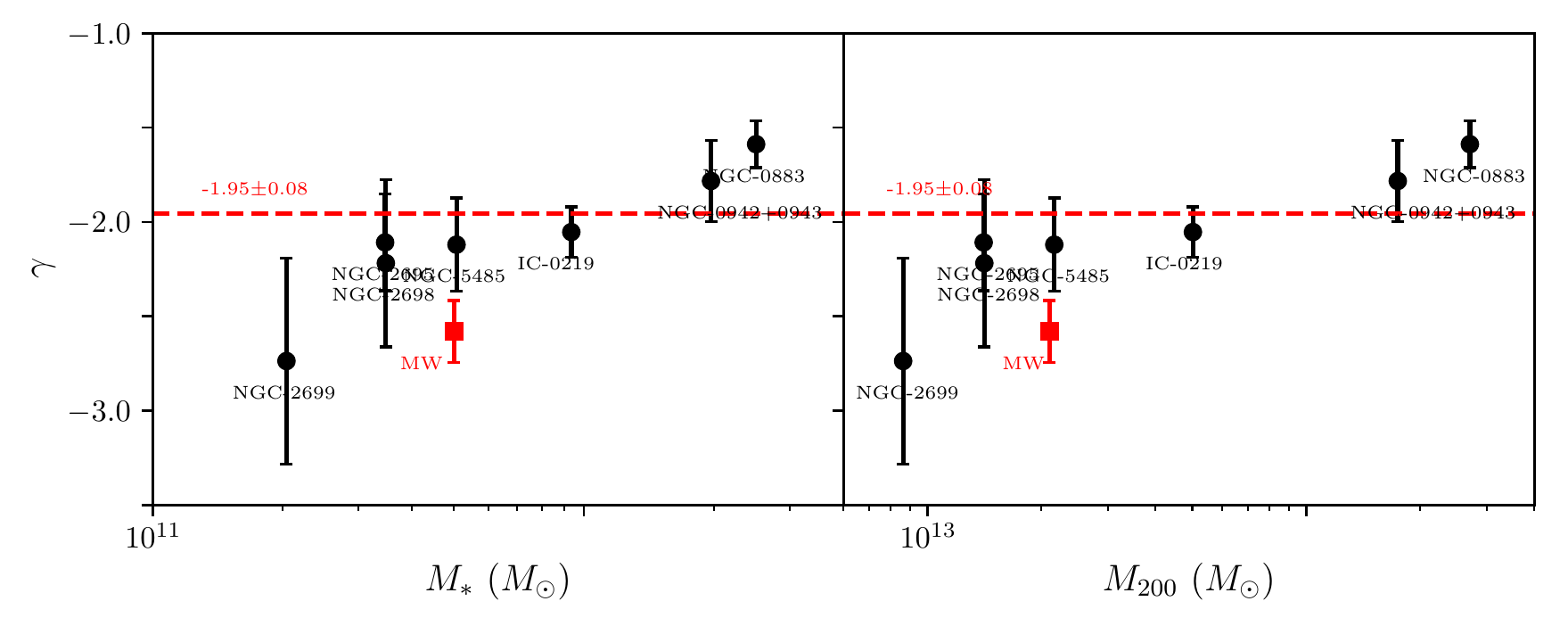}
\caption{Weighted average of the exponential term $\gamma$ in the power law fit.}
\label{fig:gammavsM}
\end{figure*}

The slope of the power law fit, $\gamma$, was plotted against $M_*$ and $M_{200}$ in \figref{gammavsM}. The weighted average of $\gamma$ was found to be $\gamma = -1.95$. The MW is significantly lower on this plots than the other galaxies. Note, however, thatr the MW power law fit included a core, whereas the power law fits of the other galaxies did not. This could possibly have an effect on the $\gamma$ of the power law. While in the previous plots there was a clear tend of increasing $R\sbr{e}$, in this plot the trend is not as convincing. \cite{Kissler97} observes a trend of increasing density profile slope with galaxy luminosity, and therefore mass. However, he argues that it is not a continuous trend, but evidence for the existence of two distinct types of GCSs.

\section{Discussion}
\label{sec:discuss}

\subsection{Scaling with halo properties}

The tight scaling of the sizes of GCSs with halo mass or virial radius suggests that the same accretion and stripping mechanisms that builds the DM halo may also build the bulk of the GCS. 
 The origin of the steep non-linear dependence of the effective radius of GCS on the virial radius of the halo ($R\sbr{e} \propto R_{200}^{3}$) is less clear. It is likely that this is a product of the hierarchical assembly of haloes in $\Lambda$CDM. Consider, for example, a $\sim L_{*}$-galaxy, like the MW or M31. As discussed above, the GCS of such a galaxy may be a combination of ``in-situ'' GCs associated with the disk and bulge, plus, as suggested by \cite{SeaZin78}, GCs associated with tidally-stripped infalling satellite galaxies.  The stripped GCs would be found at large galactocentric radii, leading or lagging the orbit of the infalling satellite. This is the case for outer halo GCs associated with the Sagittarius dwarf spheroidal \citep{BelFerIba03}.

Now consider what happens when such an $L_{*}$-galaxy merges with a galaxy group, and is itself tidally stripped. The least-bound GCs around the infalling galaxy, i.e.\ those that are most distant from it, are most easily stripped. They will become associated with the GCS of the group as a whole and their orbits will now be of order of the virial radius of the group into which they have been accreted. At the same time, some of the more tightly bound GCs may not be stripped until a later pericentric passage, at which time dynamical friction has reduced the orbit of the satellite. These GCs would be deposited at smaller group-centric radii.  Finally, when that group merges into a larger galaxy cluster, the process repeats again. In this way, a considerable fraction of the GC population may be stripped and orbit the new halo at ever increasing radii.  This provides a mechanism to explain the steep scaling relation between GCS size and halo size.

\subsection{Comparison with predictions from models}

There have been few theoretical models that have studied how the  spatial distribution of GCs develops in hierarchical models of structure formation. A notable exception is \cite{BekYahNag08}, who followed GC-like particles through a cosmological simulation. They identified DM haloes at $z \gtrsim 6$ and tagged the central particles in those haloes as GC particles. They then tracked the GC particles to the present day.  They predict a scaling of the effective radii of these metal-poor (blue) GCs with $R\sbr{e} \propto M_{200}^{0.18}$.  At face value, this is correlation is much flatter than the value found in this paper. However, their treatment of GCs is different from that adopted here. First, they only consider metal-poor/blue GCs. Second, they exclude ``intragroup'' GCs, whereas in our data any ``intragroup'' GCs are assigned to galaxies.

Most other models, whether theoretical \citep{Kru15} or semi-analytic coupled to merger trees \citep{Ton13, LiGne14}, predict abundances and metallicities, but do not predict the \emph{spatial} distribution of GCs. For example, \cite{Boy17} has proposed that the linear scaling of GC number with halo mass is largely due to a linear scaling of the blue GC number with halo mass for haloes with $M \gtrsim 10^{9} M\sun$ at $z \gtrsim 6$. At face value, this appears to be similar to the assumptions in \cite{BekYahNag08}. It would be interesting to see if the observed radial scaling is also predicted in this simple halo-based model.

\subsection{Red and blue GCs}

Most previous studies have found that the metal-poor/blue GC population is more extended than the metal-rich/red one. For the galaxies studied in this paper, the uncertainties on the individual subpopulation sizes are large and so, for most galaxies, the sizes of the two subpopulations are consistent with being equal. One outlier is NGC 2699 which has a larger red GC population than its blue population. Examination of its radial GC density profile in \figref{density} suggests that the red fit may be biased up by a ``bump'' in the red GC counts in the range $R \sim 40$--50 kpc. NGC 2699 has the poorest GCS of galaxies studied in this paper, and so it may be susceptible to contamination from the richer GCS around NGC 2698, which is $\sim 35$ kpc away on the plane of the sky. The other outlier is NGC 5485, which has a much smaller red GC $R\sbr{e}$ than that of its blue GC population.

Previous work has shown that metallicity of the red GC subpopulation is similar to the metallicity of the starlight measured the same galactocentric radius \citep{PasForUsh15}. Moreover, while we have shown that the $R\sbr{e}$ of the GCS is larger than that of the galaxy light by a factor $\sim 3$, if the red GCS  $R\sbr{e}$ is smaller then it may be close to that of the galaxy light, as it is in some galaxies \citep{Kartha16}. 

These correlations suggest a physical connection between the red GCs and the stars in the host galaxy. More specifically, this population may have been formed ``in-situ,'' with the blue GCs accreted during hierarchical assembly.  In this scenario, one might expect the effective radius of the red GCS to be more tightly linked to the galaxy light $R\sbr{e}$ while the blue GCS might scale more tightly with the halo mass or virial radius. With larger, deeper and more homogeneous samples of GCS, it should be possible to test these predictions.

\subsection{Environmental Effects}

Assuming that stripping is the dominant astrophysical process responsible for the scaling of the GCS size with halo mass, one would expect significant environmental effects. Specifically, GCs that have been stripped from satellite galaxies and orbit the host halo are then assigned to the GCS of the central galaxy in the halo. Consequently, we expect that, all other things being equal, satellite galaxies should have both lower GCS counts and masses and a smaller GCS $R\sbr{e}$, whereas the reverse should be true of centrals, particularly is they are central ellipticals. For the galaxies studied in this paper, there is a slight hint that the more massive galaxy (presumed to be the central) in the group lies somewhat higher above the mean $R\sbr{e}$-$M_{200}$ relation than subdominant (presumably satellite) galaxies.  For the galaxies collected from the literature, most of these are central galaxies in groups, or isolated field galaxies and so again, the samples are not sufficient to make this distinction. Clearly as measurements of the GCS $R\sbr{e}$ improve in quality and quantity, it will be possible to test for environmental effects.

\subsection{Future prospects}

To make significant progress in this area, a systematic multi-band survey with sufficient depth to detect most of the GC populations in nearby galaxies is required. The traditional approach has been targetted photometry of individual galaxies or galaxy clusters (e.g. NGVS). In this paper, we adopted a different approach, targetting foreground galaxies in ``blank fields'' chosen originally for weak gravitational lensing. In the current era, deep, multiband surveys covering significant fractions of the sky will  are planned or underway, including 
the Dark Energy Survey \citep{DarAbbAbd16}, 
the Canada-France Imaging Survey \citep{IbaMcCCui17} ,
the Hyper-Suprime Camera Survey \citep{AihArmBic17},
the LSST \citep{LSST09},
the Euclid mission \citep{LauAmiArd11} and
the WFIRST mission \citep{SpeGehBal15}. 
Systematic measurement of GC systems is one area that will benefit from these multi-colour panoramic surveys.

\section{Conclusions}
\label{sec:conc}

We have shown that the size of the GCS is more closely linked to the halo properties of its host dark matter halo (or, equivalently, to the total stellar mass of the central galaxy) than it is to the effective radius of the galaxy star light. The GCS size is not simply a fixed fraction of the virial radius  but rather scales steeply with the virial radius of the halo: $R\sbr{e} \propto R_{200}^{2.5-3}$. 

Dark matter haloes are built hierarchically by the accretion and tidal stripping of smaller units.  A similar hierarchical assembly of GCSs (that are increasingly less bound as one moves up the hierarchy) likely results in the steep dependence of GCS size on halo mass.
 
\section*{Acknowledgements}

We thank Bill and Gretchen Harris for useful comments on earlier versions of this paper, as well as for encouragement to submit this for publication in a timely fashion. We thank John Lucey for providing the effective radii of galaxy light in the $K-$band for select galaxies in our sample.

MH acknowledges support from an NSERC Discovery grant, and BR acknowledges support from an NSERC USRA award and support  from the University of Waterloo.

Based on observations obtained with MegaPrime/MegaCam, a joint project of CFHT and CEA/DAPNIA, at the Canada-France-Hawaii Telescope (CFHT) which is operated by the National Research Council (NRC) of Canada, the Institute National des Sciences de l'Univers of the Centre National de la Recherche Scientifique of France, and the University of Hawaii. This research used the facilities of the Canadian Astronomy Data Centre operated by the National Research Council of Canada with the support of the Canadian Space Agency, as well as the NASA/IPAC Extragalactic Database (NED) which is operated by the Jet Propulsion Laboratory, California Institute of Technology, under contract with the National Aeronautics and Space Administration.

\bibliographystyle{mn2e} 

\begin{thebibliography}{0}
\expandafter\ifx\csname natexlab\endcsname\relax\def\natexlab#1{#1}\fi

\end{thebibliography}


\begin{thebibliography}{101}
\expandafter\ifx\csname natexlab\endcsname\relax\def\natexlab#1{#1}\fi

\bibitem[{{Agnello} {et~al}\mbox{.}(2014){Agnello}, {Evans}, {Romanowsky}, \&
  {Brodie}}]{AgnEvaRom14}
{Agnello} A., {Evans} N.~W., {Romanowsky} A.~J., {Brodie} J.~P., 2014, \mnras,
  442, 3299

\bibitem[{{Aihara} {et~al}\mbox{.}(2017){Aihara}, {Armstrong}, {Bickerton},
  {Bosch}, {Coupon}, {Furusawa}, {Hayashi}, {Ikeda}, {Kamata}, {Karoji},
  {Kawanomoto}, {Koike}, {Komiyama}, {Lupton}, {Mineo}, {Miyatake}, {Miyazaki},
  {Morokuma}, {Obuchi}, {Oishi}, {Okura}, {Price}, {Takata}, {Tanaka},
  {Tanaka}, {Tanaka}, {Uchida}, {Uraguchi}, {Utsumi}, {Wang}, {Yamada},
  {Yamanoi}, {Yasuda}, {Arimoto}, {Chiba}, {Finet}, {Fujimori}, {Fujimoto},
  {Furusawa}, {Goto}, {Goulding}, {Gunn}, {Harikane}, {Hattori}, {Hayashi},
  {Helminiak}, {Higuchi}, {Hikage}, {Ho}, {Hsieh}, {Huang}, {Huang},
  {Imanishi}, {Iwata}, {Jaelani}, {Jian}, {Kashikawa}, {Katayama}, {Kojima},
  {Konno}, {Koshida}, {Leauthaud}, {Lee}, {Lin}, {Lin}, {Mandelbaum},
  {Matsuoka}, {Medezinski}, {Miyama}, {Momose}, {More}, {More}, {Mukae},
  {Murata}, {Murayama}, {Nagao}, {Nakata}, {Niikura}, {Nishizawa}, {Oguri},
  {Okabe}, {Ono}, {Onodera}, {Onoue}, {Ouchi}, {Pyo}, {Shibuya}, {Shimasaku},
  {Simet}, {Speagle}, {Spergel}, {Strauss}, {Sugahara}, {Sugiyama}, {Suto},
  {Suzuki}, {Tait}, {Takada}, {Terai}, {Toba}, {Turner}, {Uchiyama}, {Umetsu},
  {Urata}, {Usuda}, {Yeh}, \& {Yuma}}]{AihArmBic17}
{Aihara} H. {et~al.}, 2017, ArXiv e-prints

\bibitem[{{Ashman} \& {Zepf}(1998)}]{AshZep98}
{Ashman} K.~M., {Zepf} S.~E., 1998, {Globular Cluster Systems}

\bibitem[{{Balogh} {et~al}\mbox{.}(2000){Balogh}, {Navarro}, \&
  {Morris}}]{BalNavMor00}
{Balogh} M.~L., {Navarro} J.~F., {Morris} S.~L., 2000, \apj, 540, 113

\bibitem[{{Bassino} {et~al}\mbox{.}(2006){Bassino}, {Faifer}, {Forte},
  {Dirsch}, {Richtler}, {Geisler}, \& {Schuberth}}]{BasFaiFor06}
{Bassino} L.~P., {Faifer} F.~R., {Forte} J.~C., {Dirsch} B., {Richtler} T.,
  {Geisler} D., {Schuberth} Y., 2006, \aap, 451, 789

\bibitem[{{Battistini} {et~al}\mbox{.}(1993){Battistini}, {Bonoli},
  {Casavecchia}, {Ciotti}, {Federici}, \& {Fusi-Pecci}}]{Battistini93}
{Battistini} P.~L., {Bonoli} F., {Casavecchia} M., {Ciotti} L., {Federici} L.,
  {Fusi-Pecci} F., 1993, \aap, 272, 77

\bibitem[{{Beasley} {et~al}\mbox{.}(2002){Beasley}, {Baugh}, {Forbes},
  {Sharples}, \& {Frenk}}]{BeaBauFor02}
{Beasley} M.~A., {Baugh} C.~M., {Forbes} D.~A., {Sharples} R.~M., {Frenk}
  C.~S., 2002, \mnras, 333, 383

\bibitem[{{Behroozi} {et~al}\mbox{.}(2013){Behroozi}, {Wechsler}, \&
  {Conroy}}]{BehWecCon13}
{Behroozi} P.~S., {Wechsler} R.~H., {Conroy} C., 2013, \apj, 770, 57

\bibitem[{{Bekki} {et~al}\mbox{.}(2008){Bekki}, {Yahagi}, {Nagashima}, \&
  {Forbes}}]{BekYahNag08}
{Bekki} K., {Yahagi} H., {Nagashima} M., {Forbes} D.~A., 2008, \mnras, 387,
  1131

\bibitem[{{Bell} {et~al}\mbox{.}(2003){Bell}, {McIntosh}, {Katz}, \&
  {Weinberg}}]{Bell03}
{Bell} E.~F., {McIntosh} D.~H., {Katz} N., {Weinberg} M.~D., 2003, \apjs, 149,
  289

\bibitem[{{Bellazzini} {et~al}\mbox{.}(2003){Bellazzini}, {Ferraro}, \&
  {Ibata}}]{BelFerIba03}
{Bellazzini} M., {Ferraro} F.~R., {Ibata} R., 2003, \aj, 125, 188

\bibitem[{{Binney} \& {Merrifield}(1998)}]{Binney98}
{Binney} J., {Merrifield} M., 1998, {Galactic Astronomy}

\bibitem[{{Blakeslee} {et~al}\mbox{.}(1997){Blakeslee}, {Tonry}, \&
  {Metzger}}]{BlaTonMet97}
{Blakeslee} J.~P., {Tonry} J.~L., {Metzger} M.~R., 1997, \aj, 114, 482

\bibitem[{{Bland-Hawthorn} \& {Gerhard}(2016)}]{BlaGer16}
{Bland-Hawthorn} J., {Gerhard} O., 2016, \araa, 54, 529

\bibitem[{{Blom} {et~al}\mbox{.}(2012){Blom}, {Forbes}, {Brodie}, {Foster},
  {Romanowsky}, {Spitler}, \& {Strader}}]{BloForBro12}
{Blom} C., {Forbes} D.~A., {Brodie} J.~P., {Foster} C., {Romanowsky} A.~J.,
  {Spitler} L.~R., {Strader} J., 2012, \mnras, 426, 1959

\bibitem[{{Boylan-Kolchin}(2017)}]{Boy17}
{Boylan-Kolchin} M., 2017, ArXiv e-prints

\bibitem[{{Brodie} \& {Strader}(2006)}]{BroStr06}
{Brodie} J.~P., {Strader} J., 2006, \araa, 44, 193

\bibitem[{{Caldwell} \& {Romanowsky}(2016)}]{CalRom16}
{Caldwell} N., {Romanowsky} A.~J., 2016, \apj, 824, 42

\bibitem[{{Campbell} {et~al}\mbox{.}(2014){Campbell}, {Lucey}, {Colless},
  {Jones}, {Springob}, {Magoulas}, {Proctor}, {Mould}, {Read}, {Brough},
  {Jarrett}, {Merson}, {Lah}, {Beutler}, {Cluver}, \& {Parker}}]{Campbell14}
{Campbell} L.~A. {et~al.}, 2014, \mnras, 443, 1231

\bibitem[{{Cho} {et~al}\mbox{.}(2016){Cho}, {Blakeslee}, {Chies-Santos}, {Jee},
  {Jensen}, {Peng}, \& {Lee}}]{ChoBlaChi16}
{Cho} H., {Blakeslee} J.~P., {Chies-Santos} A.~L., {Jee} M.~J., {Jensen} J.~B.,
  {Peng} E.~W., {Lee} Y.-W., 2016, \apj, 822, 95

\bibitem[{{Coenda} {et~al}\mbox{.}(2009){Coenda}, {Muriel}, \&
  {Donzelli}}]{CoeMurDon09}
{Coenda} V., {Muriel} H., {Donzelli} C., 2009, \apj, 700, 1382

\bibitem[{{Cote} {et~al}\mbox{.}(1998){Cote}, {Marzke}, \&
  {West}}]{CotMarWes98}
{Cote} P., {Marzke} R.~O., {West} M.~J., 1998, \apj, 501, 554

\bibitem[{{Dark Energy Survey Collaboration} {et~al}\mbox{.}(2016){Dark Energy
  Survey Collaboration}, {Abbott}, {Abdalla}, {Aleksi{\'c}}, {Allam}, {Amara},
  {Bacon}, {Balbinot}, {Banerji}, {Bechtol}, {Benoit-L{\'e}vy}, {Bernstein},
  {Bertin}, {Blazek}, {Bonnett}, {Bridle}, {Brooks}, {Brunner}, {Buckley-Geer},
  {Burke}, {Caminha}, {Capozzi}, {Carlsen}, {Carnero-Rosell}, {Carollo},
  {Carrasco-Kind}, {Carretero}, {Castander}, {Clerkin}, {Collett}, {Conselice},
  {Crocce}, {Cunha}, {D'Andrea}, {da Costa}, {Davis}, {Desai}, {Diehl},
  {Dietrich}, {Dodelson}, {Doel}, {Drlica-Wagner}, {Estrada}, {Etherington},
  {Evrard}, {Fabbri}, {Finley}, {Flaugher}, {Foley}, {Fosalba}, {Frieman},
  {Garc{\'{\i}}a-Bellido}, {Gaztanaga}, {Gerdes}, {Giannantonio}, {Goldstein},
  {Gruen}, {Gruendl}, {Guarnieri}, {Gutierrez}, {Hartley}, {Honscheid}, {Jain},
  {James}, {Jeltema}, {Jouvel}, {Kessler}, {King}, {Kirk}, {Kron}, {Kuehn},
  {Kuropatkin}, {Lahav}, {Li}, {Lima}, {Lin}, {Maia}, {Makler}, {Manera},
  {Maraston}, {Marshall}, {Martini}, {McMahon}, {Melchior}, {Merson}, {Miller},
  {Miquel}, {Mohr}, {Morice-Atkinson}, {Naidoo}, {Neilsen}, {Nichol}, {Nord},
  {Ogando}, {Ostrovski}, {Palmese}, {Papadopoulos}, {Peiris}, {Peoples},
  {Percival}, {Plazas}, {Reed}, {Refregier}, {Romer}, {Roodman}, {Ross},
  {Rozo}, {Rykoff}, {Sadeh}, {Sako}, {S{\'a}nchez}, {Sanchez}, {Santiago},
  {Scarpine}, {Schubnell}, {Sevilla-Noarbe}, {Sheldon}, {Smith}, {Smith},
  {Soares-Santos}, {Sobreira}, {Soumagnac}, {Suchyta}, {Sullivan}, {Swanson},
  {Tarle}, {Thaler}, {Thomas}, {Thomas}, {Tucker}, {Vieira}, {Vikram},
  {Walker}, {Wechsler}, {Weller}, {Wester}, {Whiteway}, {Wilcox}, {Yanny},
  {Zhang}, \& {Zuntz}}]{DarAbbAbd16}
{Dark Energy Survey Collaboration} {et~al.}, 2016, \mnras, 460, 1270

\bibitem[{{de Vaucouleurs} {et~al}\mbox{.}(1991){de Vaucouleurs}, {de
  Vaucouleurs}, {Corwin}, {Buta}, {Paturel}, \& {Fouqu{\'e}}}]{RC3}
{de Vaucouleurs} G., {de Vaucouleurs} A., {Corwin}, Jr. H.~G., {Buta} R.~J.,
  {Paturel} G., {Fouqu{\'e}} P., 1991, {Third Reference Catalogue of Bright
  Galaxies. Volume I: Explanations and references. Volume II: Data for galaxies
  between 0$^{h}$ and 12$^{h}$. Volume III: Data for galaxies between 12$^{h}$
  and 24$^{h}$.}

\bibitem[{{Durrell} {et~al}\mbox{.}(2014){Durrell}, {C{\^o}t{\'e}}, {Peng},
  {Blakeslee}, {Ferrarese}, {Mihos}, {Puzia}, {Lan{\c c}on}, {Liu}, {Zhang},
  {Cuillandre}, {McConnachie}, {Jord{\'a}n}, {Accetta}, {Boissier}, {Boselli},
  {Courteau}, {Duc}, {Emsellem}, {Gwyn}, {Mei}, \& {Taylor}}]{Durrell14}
{Durrell} P.~R. {et~al.}, 2014, \apj, 794, 103

\bibitem[{{Escudero} {et~al}\mbox{.}(2015){Escudero}, {Faifer}, {Bassino},
  {Calder{\'o}n}, \& {Caso}}]{Escudero15}
{Escudero} C.~G., {Faifer} F.~R., {Bassino} L.~P., {Calder{\'o}n} J.~P., {Caso}
  J.~P., 2015, \mnras, 449, 612

\bibitem[{{Faifer} {et~al}\mbox{.}(2011){Faifer}, {Forte}, {Norris}, {Bridges},
  {Forbes}, {Zepf}, {Beasley}, {Gebhardt}, {Hanes}, \& {Sharples}}]{Faifer11}
{Faifer} F.~R. {et~al.}, 2011, \mnras, 416, 155

\bibitem[{{Fleming} {et~al}\mbox{.}(1995){Fleming}, {Harris}, {Pritchet}, \&
  {Hanes}}]{FleHarPri95}
{Fleming} D.~E.~B., {Harris} W.~E., {Pritchet} C.~J., {Hanes} D.~A., 1995, \aj,
  109, 1044

\bibitem[{{Forbes} {et~al}\mbox{.}(2016){Forbes}, {Alabi}, {Romanowsky},
  {Brodie}, {Strader}, {Usher}, \& {Pota}}]{ForAlaRom16}
{Forbes} D.~A., {Alabi} A., {Romanowsky} A.~J., {Brodie} J.~P., {Strader} J.,
  {Usher} C., {Pota} V., 2016, \mnras, 458, L44

\bibitem[{{Forbes} {et~al}\mbox{.}(1997){Forbes}, {Brodie}, \&
  {Grillmair}}]{Forbes97}
{Forbes} D.~A., {Brodie} J.~P., {Grillmair} C.~J., 1997, \aj, 113, 1652

\bibitem[{{Forbes} {et~al}\mbox{.}(1996){Forbes}, {Franx}, {Illingworth}, \&
  {Carollo}}]{Forbes96}
{Forbes} D.~A., {Franx} M., {Illingworth} G.~D., {Carollo} C.~M., 1996, \apj,
  467, 126

\bibitem[{{Fukugita} {et~al}\mbox{.}(1995){Fukugita}, {Shimasaku}, \&
  {Ichikawa}}]{Fukugita95}
{Fukugita} M., {Shimasaku} K., {Ichikawa} T., 1995, \pasp, 107, 945

\bibitem[{{Gill} {et~al}\mbox{.}(2005){Gill}, {Knebe}, \&
  {Gibson}}]{GilKneGib05}
{Gill} S.~P.~D., {Knebe} A., {Gibson} B.~K., 2005, \mnras, 356, 1327

\bibitem[{{Gillis} {et~al}\mbox{.}(2013){Gillis}, {Hudson}, {Erben}, {Heymans},
  {Hildebrandt}, {Hoekstra}, {Kitching}, {Mellier}, {Miller}, {van Waerbeke},
  {Bonnett}, {Coupon}, {Fu}, {Hilbert}, {Rowe}, {Schrabback}, {Semboloni}, {van
  Uitert}, \& {Velander}}]{GilHudErb13}
{Gillis} B.~R. {et~al.}, 2013, \mnras, 431, 1439

\bibitem[{{Girardi} {et~al}\mbox{.}(2003){Girardi}, {Mardirossian}, {Marinoni},
  {Mezzetti}, \& {Rigoni}}]{Girardi03}
{Girardi} M., {Mardirossian} F., {Marinoni} C., {Mezzetti} M., {Rigoni} E.,
  2003, \aap, 410, 461

\bibitem[{{Graham} \& {Driver}(2005)}]{Graham05}
{Graham} A.~W., {Driver} S.~P., 2005, \pasa, 22, 118

\bibitem[{{Gwyn}(2008)}]{Gwyn08}
{Gwyn} S.~D.~J., 2008, \pasp, 120, 212

\bibitem[{{Hargis} \& {Rhode}(2012)}]{Hargis12}
{Hargis} J.~R., {Rhode} K.~L., 2012, \aj, 144, 164

\bibitem[{{Hargis} \& {Rhode}(2014)}]{Hargis14}
{Hargis} J.~R., {Rhode} K.~L., 2014, \apj, 796, 62

\bibitem[{{Harris} {et~al}\mbox{.}(2012){Harris}, {G{\'o}mez}, {Harris},
  {Johnston}, {Kazemzadeh}, {Kerzendorf}, {Geisler}, \&
  {Woodley}}]{HarGomHar12}
{Harris} G.~L.~H., {G{\'o}mez} M., {Harris} W.~E., {Johnston} K., {Kazemzadeh}
  F., {Kerzendorf} W., {Geisler} D., {Woodley} K.~A., 2012, \aj, 143, 84

\bibitem[{{Harris}(1986)}]{Har86}
{Harris} W.~E., 1986, \aj, 91, 822

\bibitem[{{Harris}(1996)}]{Harris96}
{Harris} W.~E., 1996, \aj, 112, 1487

\bibitem[{{Harris}(2009)}]{Har09}
{Harris} W.~E., 2009, \apj, 703, 939

\bibitem[{{Harris}(2010)}]{Harris10}
{Harris} W.~E., 2010, ArXiv e-prints

\bibitem[{{Harris}(2016)}]{Har16}
{Harris} W.~E., 2016, \aj, 151, 102

\bibitem[{{Harris} {et~al}\mbox{.}(2015){Harris}, {Harris}, \&
  {Hudson}}]{Harris15}
{Harris} W.~E., {Harris} G.~L., {Hudson} M.~J., 2015, \apj, 806, 36

\bibitem[{{Harris} {et~al}\mbox{.}(2013){Harris}, {Harris}, \&
  {Alessi}}]{Harris13}
{Harris} W.~E., {Harris} G.~L.~H., {Alessi} M., 2013, \apj, 772, 82

\bibitem[{{Hudelot} {et~al}\mbox{.}(2012){Hudelot}, {Cuillandre}, {Withington},
  {Goranova}, {McCracken}, {Magnard}, {Mellier}, {Regnault}, {Betoule},
  {Aussel}, {Kavelaars}, {Fernique}, {Bonnarel}, {Ochsenbein}, \&
  {Ilbert}}]{Hudelot12}
{Hudelot} P. {et~al.}, 2012, VizieR Online Data Catalog, 2317, 0

\bibitem[{{Hudson} {et~al}\mbox{.}(2015){Hudson}, {Gillis}, {Coupon},
  {Hildebrandt}, {Erben}, {Heymans}, {Hoekstra}, {Kitching}, {Mellier},
  {Miller}, {Van Waerbeke}, {Bonnett}, {Fu}, {Kuijken}, {Rowe}, {Schrabback},
  {Semboloni}, {van Uitert}, \& {Velander}}]{Hudson15}
{Hudson} M.~J. {et~al.}, 2015, \mnras, 447, 298

\bibitem[{{Hudson} {et~al}\mbox{.}(2014){Hudson}, {Harris}, \&
  {Harris}}]{Hudson14}
{Hudson} M.~J., {Harris} G.~L., {Harris} W.~E., 2014, \apjl, 787, L5

\bibitem[{{Huxor} {et~al}\mbox{.}(2011){Huxor}, {Ferguson}, {Tanvir}, {Irwin},
  {Mackey}, {Ibata}, {Bridges}, {Chapman}, \& {Lewis}}]{HuxFerTan11}
{Huxor} A.~P. {et~al.}, 2011, \mnras, 414, 770

\bibitem[{Ibata(2017)}]{IbaMcCCui17}
Ibata R.~A., 2017

\bibitem[{{Jarrett} {et~al}\mbox{.}(2000){Jarrett}, {Chester}, {Cutri},
  {Schneider}, {Skrutskie}, \& {Huchra}}]{Jarrett00}
{Jarrett} T.~H., {Chester} T., {Cutri} R., {Schneider} S., {Skrutskie} M.,
  {Huchra} J.~P., 2000, \aj, 119, 2498

\bibitem[{{Jarrett} {et~al}\mbox{.}(2003){Jarrett}, {Chester}, {Cutri},
  {Schneider}, \& {Huchra}}]{JarCheCut03}
{Jarrett} T.~H., {Chester} T., {Cutri} R., {Schneider} S.~E., {Huchra} J.~P.,
  2003, \aj, 125, 525

\bibitem[{{Kartha} {et~al}\mbox{.}(2016){Kartha}, {Forbes}, {Alabi}, {Brodie},
  {Romanowsky}, {Strader}, {Spitler}, {Jennings}, \& {Roediger}}]{Kartha16}
{Kartha} S.~S. {et~al.}, 2016, \mnras, 458, 105

\bibitem[{{Kartha} {et~al}\mbox{.}(2014){Kartha}, {Forbes}, {Spitler},
  {Romanowsky}, {Arnold}, \& {Brodie}}]{Kartha14}
{Kartha} S.~S., {Forbes} D.~A., {Spitler} L.~R., {Romanowsky} A.~J., {Arnold}
  J.~A., {Brodie} J.~P., 2014, \mnras, 437, 273

\bibitem[{{Kissler-Patig}(1997)}]{Kissler97}
{Kissler-Patig} M., 1997, \aap, 319, 83

\bibitem[{{Ko} \& {Im}(2005)}]{KoIm05}
{Ko} J., {Im} M., 2005, Journal of Korean Astronomical Society, 38, 149

\bibitem[{{Kruijssen}(2015)}]{Kru15}
{Kruijssen} J.~M.~D., 2015, \mnras, 454, 1658

\bibitem[{{Laureijs} {et~al}\mbox{.}(2011){Laureijs}, {Amiaux}, {Arduini},
  {Augu{\`e}res}, {Brinchmann}, {Cole}, {Cropper}, {Dabin}, {Duvet}, {Ealet},
  \& et~al.}]{LauAmiArd11}
{Laureijs} R. {et~al.}, 2011, ArXiv e-prints

\bibitem[{{Lavaux} \& {Hudson}(2011)}]{Lavaux11}
{Lavaux} G., {Hudson} M.~J., 2011, \mnras, 416, 2840

\bibitem[{{Lee} {et~al}\mbox{.}(2010){Lee}, {Park}, \& {Hwang}}]{Lee10}
{Lee} M.~G., {Park} H.~S., {Hwang} H.~S., 2010, Science, 328, 334

\bibitem[{{Li} \& {Gnedin}(2014)}]{LiGne14}
{Li} H., {Gnedin} O.~Y., 2014, \apj, 796, 10

\bibitem[{{Li} {et~al}\mbox{.}(2016){Li}, {Shan}, {Kneib}, {Mo}, {Rozo},
  {Leauthaud}, {Moustakas}, {Xie}, {Erben}, {Van Waerbeke}, {Makler}, {Rykoff},
  \& {Moraes}}]{LiShaKne16}
{Li} R. {et~al.}, 2016, \mnras, 458, 2573

\bibitem[{{Li} {et~al}\mbox{.}(2014){Li}, {Shan}, {Mo}, {Kneib}, {Yang}, {Luo},
  {van den Bosch}, {Erben}, {Moraes}, \& {Makler}}]{LiShaMo14}
{Li} R. {et~al.}, 2014, \mnras, 438, 2864

\bibitem[{{Limousin} {et~al}\mbox{.}(2007){Limousin}, {Kneib}, {Bardeau},
  {Natarajan}, {Czoske}, {Smail}, {Ebeling}, \& {Smith}}]{LimKneBar07}
{Limousin} M., {Kneib} J.~P., {Bardeau} S., {Natarajan} P., {Czoske} O.,
  {Smail} I., {Ebeling} H., {Smith} G.~P., 2007, \aap, 461, 881

\bibitem[{{LSST Science Collaboration} {et~al}\mbox{.}(2009){LSST Science
  Collaboration}, {Abell}, {Allison}, {Anderson}, {Andrew}, {Angel}, {Armus},
  {Arnett}, {Asztalos}, {Axelrod}, \& et~al.}]{LSST09}
{LSST Science Collaboration} {et~al.}, 2009, ArXiv e-prints

\bibitem[{{Ludlow} {et~al}\mbox{.}(2009){Ludlow}, {Navarro}, {Springel},
  {Jenkins}, {Frenk}, \& {Helmi}}]{LudNavSpr09}
{Ludlow} A.~D., {Navarro} J.~F., {Springel} V., {Jenkins} A., {Frenk} C.~S.,
  {Helmi} A., 2009, \apj, 692, 931

\bibitem[{{Mackey} {et~al}\mbox{.}(2016){Mackey}, {Beasley}, \&
  {Leaman}}]{MacBeaLea16}
{Mackey} A.~D., {Beasley} M.~A., {Leaman} R., 2016, \mnras, 460, L114

\bibitem[{{Mackey} {et~al}\mbox{.}(2010){Mackey}, {Huxor}, {Ferguson}, {Irwin},
  {Tanvir}, {McConnachie}, {Ibata}, {Chapman}, \& {Lewis}}]{MacHuxFer10}
{Mackey} A.~D. {et~al.}, 2010, \apjl, 717, L11

\bibitem[{{Marinoni} \& {Hudson}(2002)}]{MarHud02}
{Marinoni} C., {Hudson} M.~J., 2002, \apj, 569, 101

\bibitem[{{Natarajan} {et~al}\mbox{.}(2009){Natarajan}, {Kneib}, {Smail},
  {Treu}, {Ellis}, {Moran}, {Limousin}, \& {Czoske}}]{NatKneSma09}
{Natarajan} P., {Kneib} J.-P., {Smail} I., {Treu} T., {Ellis} R., {Moran} S.,
  {Limousin} M., {Czoske} O., 2009, \apj, 693, 970

\bibitem[{Oman {et~al}\mbox{.}(2013)Oman, Hudson, \& Behroozi}]{OmaHudBeh13}
Oman K.~A., Hudson M.~J., Behroozi P.~S., 2013, \mnras, 431, 2307

\bibitem[{{Pastorello} {et~al}\mbox{.}(2015){Pastorello}, {Forbes}, {Usher},
  {Brodie}, {Romanowsky}, {Strader}, {Spitler}, {Alabi}, {Foster}, {Jennings},
  {Kartha}, \& {Pota}}]{PasForUsh15}
{Pastorello} N. {et~al.}, 2015, \mnras, 451, 2625

\bibitem[{{Peng} {et~al}\mbox{.}(2011){Peng}, {Ferguson}, {Goudfrooij},
  {Hammer}, {Lucey}, {Marzke}, {Puzia}, {Carter}, {Balcells}, {Bridges},
  {Chiboucas}, {del Burgo}, {Graham}, {Guzm{\'a}n}, {Hudson}, {Matkovi{\'c}},
  {Merritt}, {Miller}, {Mouhcine}, {Phillipps}, {Sharples}, {Smith}, {Tully},
  \& {Verdoes Kleijn}}]{Peng11}
{Peng} E.~W. {et~al.}, 2011, \apj, 730, 23

\bibitem[{Peng {et~al}\mbox{.}(2008)Peng, Jord{\'a}n, C{\^o}t{\'e}, Takamiya,
  West, Blakeslee, Chen, Ferrarese, Mei, Tonry, \& West}]{PenJorCot08}
Peng E.~W. {et~al.}, 2008, The Astrophysical Journal, 681, 197

\bibitem[{{Pota} {et~al}\mbox{.}(2013){Pota}, {Forbes}, {Romanowsky}, {Brodie},
  {Spitler}, {Strader}, {Foster}, {Arnold}, {Benson}, {Blom}, {Hargis},
  {Rhode}, \& {Usher}}]{PotForRom13}
{Pota} V. {et~al.}, 2013, \mnras, 428, 389

\bibitem[{{Ramos} {et~al}\mbox{.}(2015){Ramos}, {Coenda}, {Muriel}, \&
  {Abadi}}]{RamCoeMur15}
{Ramos} F., {Coenda} V., {Muriel} H., {Abadi} M., 2015, \apj, 806, 242

\bibitem[{{Rejkuba} {et~al}\mbox{.}(2014){Rejkuba}, {Harris}, {Greggio},
  {Harris}, {Jerjen}, \& {Gonzalez}}]{RejHarGre14}
{Rejkuba} M., {Harris} W.~E., {Greggio} L., {Harris} G.~L.~H., {Jerjen} H.,
  {Gonzalez} O.~A., 2014, \apjl, 791, L2

\bibitem[{{Rhode} \& {Zepf}(2004)}]{Rhode04}
{Rhode} K.~L., {Zepf} S.~E., 2004, \aj, 127, 302

\bibitem[{{Rhode} {et~al}\mbox{.}(2007){Rhode}, {Zepf}, {Kundu}, \&
  {Larner}}]{RhoZepKun07}
{Rhode} K.~L., {Zepf} S.~E., {Kundu} A., {Larner} A.~N., 2007, \aj, 134, 1403

\bibitem[{{Romanowsky} {et~al}\mbox{.}(2012){Romanowsky}, {Strader}, {Brodie},
  {Mihos}, {Spitler}, {Forbes}, {Foster}, \& {Arnold}}]{RomStrBro12}
{Romanowsky} A.~J., {Strader} J., {Brodie} J.~P., {Mihos} J.~C., {Spitler}
  L.~R., {Forbes} D.~A., {Foster} C., {Arnold} J.~A., 2012, \apj, 748, 29

\bibitem[{{Salinas} {et~al}\mbox{.}(2015){Salinas}, {Alabi}, {Richtler}, \&
  {Lane}}]{Salinas15}
{Salinas} R., {Alabi} A., {Richtler} T., {Lane} R.~R., 2015, \aap, 577, A59

\bibitem[{{Schlafly} \& {Finkbeiner}(2011)}]{Schlafly11}
{Schlafly} E.~F., {Finkbeiner} D.~P., 2011, \apj, 737, 103

\bibitem[{{Searle} \& {Zinn}(1978)}]{SeaZin78}
{Searle} L., {Zinn} R., 1978, \apj, 225, 357

\bibitem[{{S{\'e}rsic}(1963)}]{Sersic63}
{S{\'e}rsic} J.~L., 1963, Boletin de la Asociacion Argentina de Astronomia La
  Plata Argentina, 6, 41

\bibitem[{{Sersic}(1968)}]{Sersic68}
{Sersic} J.~L., 1968, {Atlas de galaxias australes}

\bibitem[{{Sick} {et~al}\mbox{.}(2015){Sick}, {Courteau}, {Cuillandre},
  {Dalcanton}, {de Jong}, {McDonald}, {Simard}, \& {Tully}}]{Sick15}
{Sick} J., {Courteau} S., {Cuillandre} J.-C., {Dalcanton} J., {de Jong} R.,
  {McDonald} M., {Simard} D., {Tully} R.~B., 2015, in IAU Symposium, Vol. 311,
  IAU Symposium, {Cappellari} M., {Courteau} S., eds., pp. 82--85

\bibitem[{{Smith} {et~al}\mbox{.}(2015){Smith}, {S{\'a}nchez-Janssen},
  {Beasley}, {Candlish}, {Gibson}, {Puzia}, {Janz}, {Knebe}, {Aguerri},
  {Lisker}, {Hensler}, {Fellhauer}, {Ferrarese}, \& {Yi}}]{SmiSanBea15}
{Smith} R. {et~al.}, 2015, \mnras, 454, 2502

\bibitem[{{Smith} {et~al}\mbox{.}(2013){Smith}, {S{\'a}nchez-Janssen},
  {Fellhauer}, {Puzia}, {Aguerri}, \& {Farias}}]{SmiSanFel13}
{Smith} R., {S{\'a}nchez-Janssen} R., {Fellhauer} M., {Puzia} T.~H., {Aguerri}
  J.~A.~L., {Farias} J.~P., 2013, \mnras, 429, 1066

\bibitem[{{Spergel} {et~al}\mbox{.}(2015){Spergel}, {Gehrels}, {Baltay},
  {Bennett}, {Breckinridge}, {Donahue}, {Dressler}, {Gaudi}, {Greene}, {Guyon},
  {Hirata}, {Kalirai}, {Kasdin}, {Macintosh}, {Moos}, {Perlmutter}, {Postman},
  {Rauscher}, {Rhodes}, {Wang}, {Weinberg}, {Benford}, {Hudson}, {Jeong},
  {Mellier}, {Traub}, {Yamada}, {Capak}, {Colbert}, {Masters}, {Penny},
  {Savransky}, {Sterns}, {Zimmerman}, {Barry}, {Bartusek}, {Carpenter},
  {Cheng}, {Content}, {Dekens}, {Demers}, {Grady}, {Jackson}, {Kuan}, {Kruk},
  {Melton}, {Nemati}, {Parvin}, {Poberezhskiy}, {Peddie}, {Ruffa}, {Wallace},
  {Whipple}, {Wollack}, \& {Zhao}}]{SpeGehBal15}
{Spergel} D. {et~al.}, 2015, ArXiv e-prints

\bibitem[{{Spitler} \& {Forbes}(2009)}]{SpiFor09}
{Spitler} L.~R., {Forbes} D.~A., 2009, \mnras, 392, L1

\bibitem[{{Strader} {et~al}\mbox{.}(2005){Strader}, {Brodie}, {Cenarro},
  {Beasley}, \& {Forbes}}]{Strader05}
{Strader} J., {Brodie} J.~P., {Cenarro} A.~J., {Beasley} M.~A., {Forbes} D.~A.,
  2005, \aj, 130, 1315

\bibitem[{{Tonini}(2013)}]{Ton13}
{Tonini} C., 2013, \apj, 762, 39

\bibitem[{{Villegas} {et~al}\mbox{.}(2010){Villegas}, {Jord{\'a}n}, {Peng},
  {Blakeslee}, {C{\^o}t{\'e}}, {Ferrarese}, {Kissler-Patig}, {Mei}, {Infante},
  {Tonry}, \& {West}}]{Villegas10}
{Villegas} D. {et~al.}, 2010, \apj, 717, 603

\bibitem[{{Voggel} {et~al}\mbox{.}(2016){Voggel}, {Hilker}, \&
  {Richtler}}]{VogHilRic16}
{Voggel} K., {Hilker} M., {Richtler} T., 2016, \aap, 586, A102

\bibitem[{{Wehner} {et~al}\mbox{.}(2008){Wehner}, {Harris}, {Whitmore},
  {Rothberg}, \& {Woodley}}]{WehHarWhi08}
{Wehner} E.~M.~H., {Harris} W.~E., {Whitmore} B.~C., {Rothberg} B., {Woodley}
  K.~A., 2008, \apj, 681, 1233

\bibitem[{{West} {et~al}\mbox{.}(1995){West}, {Cote}, {Jones}, {Forman}, \&
  {Marzke}}]{WesCotJon95}
{West} M.~J., {Cote} P., {Jones} C., {Forman} W., {Marzke} R.~O., 1995, \apjl,
  453, L77

\bibitem[{{Worthey}(1994)}]{Worthey94}
{Worthey} G., 1994, \apjs, 95, 107

\bibitem[{{Yahagi} \& {Bekki}(2005)}]{YahBek05}
{Yahagi} H., {Bekki} K., 2005, \mnras, 364, L86

\bibitem[{{Young} {et~al}\mbox{.}(2012){Young}, {Dowell}, \& {Rhode}}]{Young12}
{Young} M.~D., {Dowell} J.~L., {Rhode} K.~L., 2012, \aj, 144, 103

\end{thebibliography}

\appendix

\section{Corrections}

\subsection{GCLF, photometry and completeness corrections}
\label{sec:gclf}

To make corrections for the incompleteness of the observed GCs, we adopt the Gaussian GCLF of \cite{Villegas10}:
\begin{equation}
\label{eq:gaussian}
\frac{dN}{dm} = \frac{1}{\sqrt{2\pi\sigma}}\textrm{exp}\!\left[-\frac{(m-\mu)^2}{2\sigma^2}\right] \,
\end{equation}
where the peak ($\mu$) and standard deviation ($\sigma$) are functions of the $z$-band magnitude of the parent galaxy:
\begin{equation}
\label{eq:sigma}
\sigma_z = (1.07 \pm 0.02) - (0.10 \pm 0.01)(M_{z,gal}+22)
\end{equation}%
\begin{equation}
\label{eq:peak}
\mu_z = (-7.66 \pm 0.18) + (0.04 \pm 0.01)(M_{z,gal} +22) \,.
\end{equation}
The above equations are based on magnitudes in the F850LP ($\approx$~Sloan~$z$) passband of the Advanced Camera for Surveys (ACS) on the Hubble Space Telescope. 
In order to convert our galaxy $K$ magnitudes into Sloan $z$ magnitudes, we use average galaxy colours given in \cite{Girardi03}, adopting an average $B-K$ colour of 3.8, and \cite{Fukugita95}. Using these $z$ magnitudes, we obtain the peak and deviation of the GCLF for each galaxy.

Next, we convert the GCLF peak magnitude from SDSS $z$ to SDSS $i$. \cite{Strader05} give equations for average $g-z$ colours for red and blue GCs. Averaging these equations will yield an average $g-z$ colour for GCs. \cite{Durrell14} define GCs as existing within a CFHT $g-i$ colour range of 0.55~\textless~$g-i$~\textless~1.15. We will define the average $g-i$ of GCs as the centre of this range at $g-i$ = 0.85. To convert this colour from the CFHT filter set to the ACS filter set we use the CFHT to SDSS reverse transformations.\footnote{http://www.cadc-ccda.hia-iha.nrc-cnrc.gc.ca/en/megapipe/docs/filt.html} Using the average $g-z$ and $g-i$ colours, we convert the GCLF from SDSS $z$ to SDSS $i$.

We need to convert the GCLF from SDSS $i$ to CFHT $i$. Using the models of \cite{Worthey94}\footnote{http://astro.wsu.edu/dial/dial\_a\_model.html}, we determine the average $r-i$ colour of a sample of an equal amount of red and blue GCs. Blue GCs were treated as having an age of 12\,Gyr and [Fe/H] = -1.5. Red GCs were treated as having an age of 10\,Gyr and [Fe/H] $= -0.5$. The average $r-i$ colour of a GC was calculated as 0.267. We can use this colour and the transformations from \cite{Gwyn08} to transform our GCLF model from the ACS/SDSS filter system to the CFHT filter system. 

Galaxy and GC magnitudes are corrected for extinction using \cite{Schlafly11}.

\makeatletter{}\begin{table*}
\section{De Vaucouleurs and Power Law Fit Parameters}
\label{sec:parameters}
\caption{The fit parameters of fixed R$\sbr{e}$ de Vaucouleurs fits for galaxies studied in this paper, as well as the $\chi^2$ of the fit. In each fit there are 14 degrees of freedom.}
\begin{tabular}{llcccccc}
\hline
Galaxy&Field&Total $\Sigma\sbr{e}$&Total $\chi^2$&Red $\Sigma\sbr{e}$&Red $\chi^2$&Blue $\Sigma\sbr{e}$&Blue $\chi^2$\\
&&$(10^{-2}\frac{GC}{kpc^2})$&&$(10^{-2}\frac{GC}{kpc^2})$&&$(10^{-2}\frac{GC}{kpc^2})$&\\
\hline
IC 219&W1-0-0&25.67$\pm$2.99&6.44&12.49$\pm$2.17&6.78&10.65$\pm$2.27&8.72\\
NGC 883&W1-0-0&4.85$\pm$0.80&7.67&2.68$\pm$0.67&9.95&1.27$\pm$0.44&6.58\\
NGC 942+943&W1+3-4&7.25$\pm$1.47&16.76&2.98$\pm$0.88&13.14&3.64$\pm$0.89&13.56\\
NGC 2695&W2-0+1&4.01$\pm$0.70&11.17&2.56$\pm$0.69&16.22&0.95$\pm$0.54&18.04\\
NGC 2698&W2-0+1&2.80$\pm$0.63&9.92&1.04$\pm$0.36&7.63&1.44$\pm$0.44&9.61\\
NGC 2699&W2-0+1&1.70$\pm$0.69&7.98&1.08$\pm$0.56&8.76&1.14$\pm$0.46&6.51\\
NGC 5473&W3-2-0&2.57$\pm$0.83&19.80&0.80$\pm$0.55&20.61&1.61$\pm$0.33&5.99\\
NGC 5475&W3-2+1&1.01$\pm$0.47&7.77&1.44$\pm$0.70&6.19&0.97$\pm$0.46&8.46\\
NGC 5485&W3-2-0&7.90$\pm$1.21&11.47&2.25$\pm$0.67&8.20&5.00$\pm$1.02&13.62\\
\end{tabular}
\vspace{0.5cm}

\vspace{0.5cm}
\caption{The fit parameters of power law fits for galaxies studied in this paper, as well as the $\chi^2$ of the fit. In each fit there are 13 degrees of freedom.}
\begin{tabular}{llccccccccc}
\hline
Galaxy&Field&Total $\Sigma_0$&Total $\gamma$&Total $\chi^2$&Red $\Sigma_0$&Red $\gamma$&Red $\chi^2$&Blue $\Sigma_0$&Blue $\gamma$&Blue $\chi^2$\\
&&$(10^{-6}\frac{GC}{kpc^2})$&&&$(10^{-6}\frac{GC}{kpc^2})$&&&$(10^{-6}\frac{GC}{kpc^2})$&&\\
\hline
IC 219&W1-0-0&6450$\pm$1305&-2.07$\pm$0.14&5.53&2695$\pm$956&-2.20$\pm$0.24&6.05&2246$\pm$1030&-2.17$\pm$0.29&9.14\\
NGC 883&W1-0-0&5837$\pm$1070&-1.72$\pm$0.16&5.27&3484$\pm$999&-1.67$\pm$0.26&8.02&1482$\pm$735&-1.77$\pm$0.41&6.18\\
NGC 942+943&W1+3-4&5311$\pm$1437&-1.87$\pm$0.24&15.36&2433$\pm$958&-1.78$\pm$0.35&12.44&2631$\pm$791&-1.89$\pm$0.28&12.18\\
NGC 2695&W2-0+1&399$\pm$216&-2.10$\pm$0.26&10.64&143$\pm$138&-2.39$\pm$0.43&15.64&24$\pm$59&-2.78$\pm$1.01&17.22\\
NGC 2698&W2-0+1&228$\pm$218&-2.21$\pm$0.44&9.81&48$\pm$89&-2.49$\pm$0.82&7.32&57$\pm$84&-2.55$\pm$0.65&9.42\\
NGC 2699&W2-0+1&28$\pm$44&-2.73$\pm$0.54&6.58&15$\pm$35&-2.77$\pm$0.77&7.92&157$\pm$180&-1.81$\pm$0.48&6.31\\
NGC 5473&W3-2-0&54$\pm$127&-3.01$\pm$1.03&18.54&0$\pm$1&-5.00$\pm$2.87&18.83&104$\pm$119&-2.49$\pm$0.53&5.75\\
NGC 5475&W3-2+1&219$\pm$224&-1.58$\pm$0.56&7.14&26$\pm$47&-2.77$\pm$0.75&5.73&99$\pm$139&-1.92$\pm$0.79&8.26\\
NGC 5485&W3-2-0&1056$\pm$506&-2.12$\pm$0.25&10.73&452$\pm$363&-1.92$\pm$0.44&7.48&375$\pm$263&-2.43$\pm$0.35&12.88\\
\label{tab:fixparam}
\end{tabular}
\end{table*}

\end{document}